\documentclass[twocolumn]{aastex631}

\usepackage{graphicx}
\usepackage{array}
\usepackage{xcolor}

\received{March 28, 2024}
\revised{May 23, 2024}
\accepted{June 7 2024}

\shorttitle{Sparse Photometry of JFCs in the LSST Era}
\shortauthors{Donaldson et al.}

\begin{document}

\title{Predictions for Sparse Photometry of Jupiter-family Comet Nuclei in the LSST Era}


\correspondingauthor{A. Donaldson}
\email{a.donaldson@ed.ac.uk}

\author[0000-0003-4507-9384]{A. Donaldson}
\affiliation{Institute for Astronomy, University of Edinburgh, EH9 3HJ, Edinburgh, UK}

\author{C. Snodgrass}
\affiliation{Institute for Astronomy, University of Edinburgh, EH9 3HJ, Edinburgh, UK}

\author{R. Kokotanekova}
\affiliation{Institute of Astronomy and National Astronomical Observatory, Bulgarian Academy of Sciences, 72 Tsarigradsko Chaussee Blvd., 1784 Sofia, Bulgaria}

\author{A. Rożek}
\affiliation{Institute for Astronomy, University of Edinburgh, EH9 3HJ, Edinburgh, UK}



\begin{abstract}

The Legacy Survey of Space and Time (LSST) at Vera C. Rubin Observatory will deliver high-quality, temporally-sparse observations of millions of Solar System objects on an unprecedented scale. Such datasets will likely enable the precise estimation of small body properties on a population-wide basis. In this work, we consider the possible applications of photometric data points from LSST to the characterisation of Jupiter-family comet (JFC) nuclei. 
We simulate sparse-in-time lightcurve points with an LSST-like cadence for the orbit of a JFC between 2024-2033. 
Convex lightcurve inversion is used to assess whether the simulation input parameters can be accurately reproduced for a sample of nucleus rotation periods, pole orientations, activity onsets, shapes and sizes.
We find that the rotation period and pole direction can be reliably constrained across all nucleus variants tested,
and that the convex shape models, while limited in their ability to describe complex or bilobed nuclei, are effective for correcting sparse photometry for rotational modulation to  
improve estimates of nucleus phase functions.
Based on this analysis, we anticipate that LSST photometry will significantly enhance our present understanding of the spin-state and phase function distributions of JFC nuclei.

\end{abstract}



\section{Introduction} \label{sec:intro}

Much of our present understanding of comet physical properties has been derived from observations of their nuclei in broad-band optical filters. Jupiter-family comets (JFCs) are an intriguing and informative population of comets to study in this way. The source populations of these objects lie beyond the orbit of Neptune in dynamically unstable regions of the Kuiper Belt before being propelled to low-inclination, short-period ($<20$ years) orbits following gravitational interactions with the giant planets \citep{levison_kuiper_1997}. Studying the physical characteristics of JFCs provides insights into the processes they experience during their time in the inner Solar System, which is essential for contextualising their evolution from their reservoirs of origin.

 As comet nuclei are km-sized, these objects typically remain observable by 1-4m class observing facilities over most of their orbits. Near the Sun, the temperature is high enough that their surfaces are shrouded to the ground-based observer by a diffuse coma of gas and dust driven by the sublimation of volatile species in the nucleus. For most JFCs, sublimative activity diminishes with increased heliocentric distance, at which point the dominant signal from the comet becomes sunlight reflected from the nucleus surface. When this is the case, the ground-based observer can estimate nucleus properties depending on the amount and quality of available data. As a starting point, multi-band snapshot photometry allows for the absolute magnitude of the nucleus to be measured in filters of interest, placing constraints on the colour and effective size of the nucleus \citep[e.g.][]{licandro_ccd_2000,lowry_ccd_2003}. This technique is somewhat limited in that it captures nucleus information at only a single instance of its rotational phase. With sufficiently long sequences of dense-in-time imaging, typically in a single filter, the rotation period of the nucleus about its spin axis can be measured. This estimate improves in reliability with lightcurves collected across successive nights, probing larger portions of the object's rotational phase without notably altering the observing geometry of the nucleus with respect to the observer. 
 
 From lightcurves collected across multiple epochs spanning a range of solar phase angles, the dependence of nucleus brightness on phase angle can be estimated \citep[e.g.][]{lamy_sizes_2004, snodgrass_size_2011, kokotanekova_rotation_2017}. Combined with reliable nucleus size measurements, the resulting phase function allows for the geometric albedo ($p_V$) of the nucleus to be estimated. With a large enough number of observational epochs, it is possible to fit a convex model to the collection of lightcurves to identify a likely shape and pole orientation for the nucleus \citep{kaasalainen_optimization_2001-1, kaasalainen_optimization_2001}. This was recently demonstrated by \cite{donaldson_characterizing_2023} for a JFC, using absolutely-calibrated dense-in-time lightcurves of 162P/Siding Spring spanning more than fifteen years of observations. The resulting shape model indicated an elongated shape for the nucleus of 162P and was used to correct the phase function for rotational variation. 
 A similar approach was used by \citet{lowry_nucleus_2012} and \citet{mottola_rotation_2014} to accurately predict the spin state of JFC 67P/Churyumov-Gerasimenko prior to the arrival of the {\em Rosetta} spacecraft.
 Using lightcurve inversion to model the shape of a comet nucleus provides the most detailed information possible from the ground on its shape and spin state. Such comprehensive information for a greater number of JFCs may provide insights into the distribution of these properties for a more representative sample of the population and offer new perspectives into the mechanisms driving their activity behaviours and evolutionary processes. 

Given the limited number of comets for which large amounts of reliable, dense-in-time lightcurve data are presently available, we consider the role that large-scale optical surveys might play in enhancing the determination of JFC nucleus properties. Optical surveys can obtain intermittent photometry of small bodies contained within their field of view - often fortuitously - with the regularity of data points for each object observed dependent on survey cadence and orbital parameters. In recent decades, survey telescopes have become increasingly powerful, reaching fainter limiting magnitudes and covering greater areas of sky, resulting in the discovery of thousands more small bodies. 
\cite{kaasalainen_physical_2004} showed it was theoretically possible to obtain reliable asteroid shape and spin estimates from solely sparse-in-time photometry. Data from surveys such as the Asteroid Terrestrial-impact Last Alert System (ATLAS) and Zwicky Transient Facility (ZTF) have produced reliable shape and period estimates of asteroid populations solely using sparse photometry over long temporal baselines \citep[e.g.][]{durech_asteroid_2020, durech_reconstruction_2023}. 

Such observations of comet nuclei are typically more challenging. To avoid contamination from activity in photometry apertures, we are restricted to observing them at large heliocentric distances (typically $>3$ au) where they are faintest. Small survey telescopes are therefore unlikely to detect inactive comet nuclei at sufficient signal-to-noise for reliable photometry \citep[for a recent summary of the contributions of telescopic surveys to comet studies, see][]{bauer_comet_2022}.

The Legacy Survey of Space and Time (LSST) at the Vera C. Rubin Observatory on Cerro Pachón, Chile, will survey 18,000 deg$^2$ of the southern sky in six $ugrizy$ filters over its 10-year period of operation \citep{ivezic_lsst_2019, bianco_f_survey_2022}. At the time of writing, the survey is expected to commence in mid-2025. The primary observing mode of the LSST is a `wide-fast-deep' (WFD) optical survey covering the entire observable sky in 
approximately
three-day intervals. Additional time will be allocated to several micro-surveys probing smaller sky regions to greater depth. The survey will be conducted with the Simonyi Survey Telescope, which has a primary mirror diameter of 8.4-metres. The design is such that the 
single exposure
limiting object brightness in the $r$-band for the WFD mode is around 24th magnitude.
It is currently predicted that the survey will discover more than five million new solar system objects, and will provide unprecedented numbers of calibrated photometric data points for the moving objects it detects. The large survey area coupled with the faint limiting magnitude have profound implications for the study of the rotational lightcurves of faint small bodies. LSST will likely be the first survey capable of reliably acquiring sparse photometry of inactive short period comets. Given the challenges presented by observing comet nuclei at a wide range of observing geometries with targeted observations, will nucleus properties such as pole orientation and shape be accessible with LSST?

The LSST science priorities encompass 
many astronomical disciplines in addition to inventorying the Solar System. 
Finding an observing strategy that will optimise the scientific returns in each of these areas is not straightforward. The intricacies of the survey cadence are therefore not yet finalised, and will likely continue to be tweaked throughout the early years of the LSST. 
This work does not aim to contribute to the discussion on survey cadence specifics, and 
the methods and results discussed here by no means reflect the exact final output from the LSST. 
Our goal is to 
examine the potential of sparse-in-time LSST photometry for the characterisation of JFC nuclei.
To date, no survey has been capable of obtaining high-quality nucleus photometry on such vast scales, and our focus is the application of this photometry to measuring nucleus properties on an individual object basis.


In this work we simulate observations of typical JFC nuclei with an LSST-like cadence.
Section \ref{synth} describes the process for generating synthetic sparse-in-time nucleus lightcurves using existing survey simulator tools. 
Section \ref{analysis} details the lightcurve analysis methods we employ by application to a noiseless example case. 
By individually varying the input parameters of the nucleus model, we examine how intrinsic nucleus properties affect the outcomes of lightcurve analysis in Section \ref{params}. 
In Section \ref{mc} we incorporate the effects of photometric uncertainty on the synthetic lightcurve points and analyse these using a Monte Carlo approach. We also evaluate combining the sparse survey data points with simulated dense-in-time lightcurves.
Finally in Section \ref{disc} we assess our findings in a realistic context beyond the simplifying assumptions made to generate the lightcurve datasets, and discuss their implications for how reliably we can expect to measure JFC properties using LSST photometry.

\section{Generating synthetic nucleus lightcurves}
\label{synth}

\subsection{Simulating lightcurve points with an LSST-like cadence}
We first consider the temporal spacing of the synthetic lightcurve points. The timestamps i.e. when the comet is contained within the LSST field of view (FOV), depend on both the observing cadence of the survey and the orbital parameters of the comet. To create these timestamps, we use the \texttt{objectsInField} (OIF) module within the Asteroid Survey Simulator package \citep{cornwall_simulated_2020}. For an input population of small bodies, OIF identifies field pointings in which the specified objects are detected. We chose the orbit of comet 67P/Churyumov-Gerasimenko (hereafter 67P) as it is representative of a `typical' JFC orbit and is not expected to undergo any significant orbital modifications during the LSST operation period. We used a polygonal survey FOV with an effective area of 10.5 deg$^2$ to simulate the array of pixel sensors on the LSST Camera \cite[Fig. 12]{ivezic_lsst_2019}. The survey cadence used was \textit{baseline\textunderscore3.0}, the most current database of simulated cadence information available at the time. The database contains simulated observational properties of each pointing such as filter, visit start time, and the limiting magnitude at that pointing. It should be noted that \textit{baseline\textunderscore3.0} is based on the recommendations of the Survey Cadence Optimization Committee in December 2022 and does not reflect the exact final observing strategy for LSST - rather, it incorporates updates from previous versions of the observing strategy based on feedback \citep{bianco_f_survey_2022}. 

The resulting dataset from which we construct the nucleus lightcurves consists of 613 individual field pointings between January 2024 and September 2033.  The median effective seeing FWHM is 1.03 arcseconds for these pointings, and throughout this work we assume that the object's proper motion is sufficiently slow that it would not be trailed in the $\sim$30-second LSST exposures. In the \textit{baseline\textunderscore3.0} cadence, most of the $grizy$ pointings are comprised of two 15-second exposures in each filter, separated by a shutter-close time of 1 second and a readout time of 2 seconds. To define the lightcurve timestamps, we take the midpoint of these pointings to be 16.5 seconds after the start time of the field pointing. Some of the $r$- and $i$-band exposures are single 15-second exposures, for which we take the timestamp to be 7.5 seconds after the pointing start time. The $u$-band pointings in the \textit{baseline\textunderscore3.0} cadence are entirely single 30-second visits, chosen to minimise read noise in this filter. We take the midpoint as being 15s from the start time of the pointing. We correct each timestamp for light-travel time using the geocentric distance of the comet so that each timestamp depicts the time at which the light left the nucleus surface at the exposure midpoint.

\subsection{Nucleus magnitudes}
\label{sim_mags}

\begin{figure*}
\centering
\noindent
\begin{tabular}{c c}
        \includegraphics[width=3in]{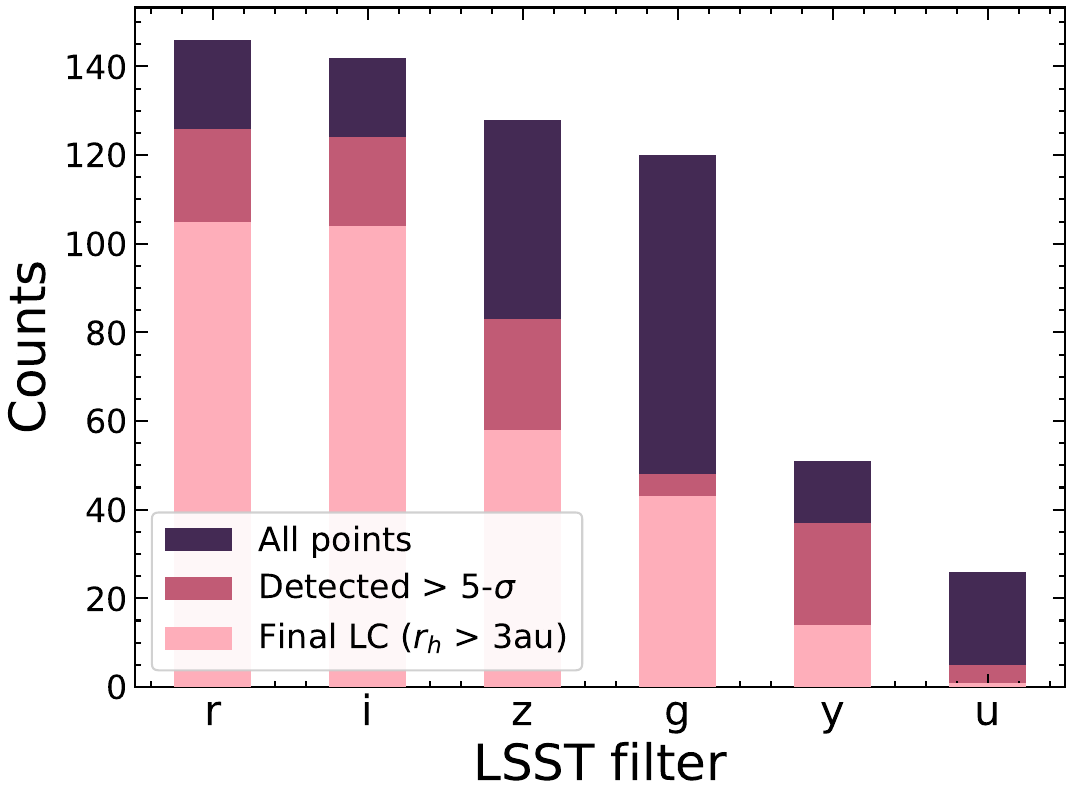}&
            \includegraphics[width=3.75in]{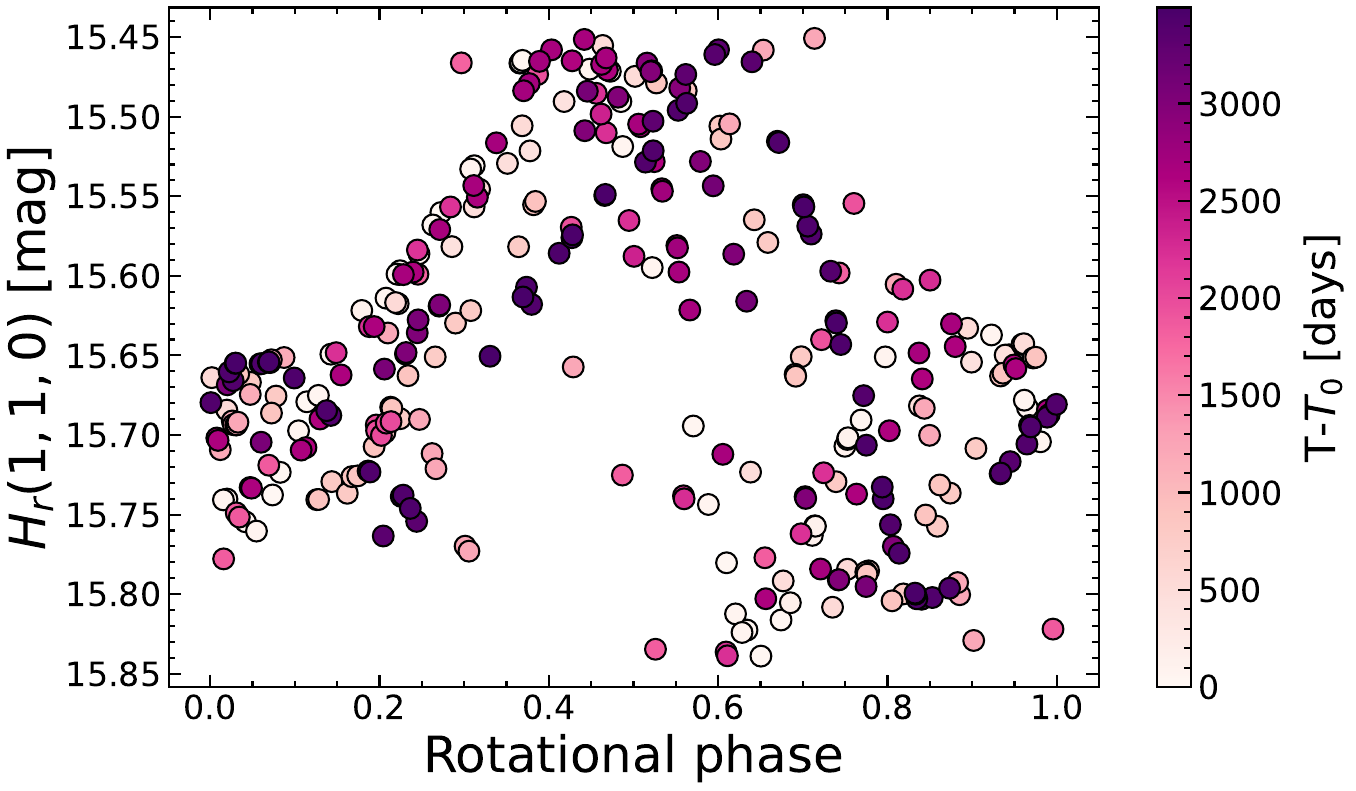} \\
            (a) & (b) \\
\end{tabular}
\caption{Left: filter (LSST $ugrizy$) of every pointing covered by the simulated nominal dataset. Lightcurve points are discarded from the final dataset first by SNR, then heliocentric distance $r_h$. Right: final lightcurve for the nominal case of this study, with 325 individual points remaining within imposed 5$\sigma$ and $r_h$ limits. Marker colour represents time from first observation. Points were corrected for solar phase by linear phase function slope value $\beta_L$, and phased to period $P_i=14.8978$~h. The misalignment of lightcurve profiles is due to synodic effects (i.e. the object viewing geometry varying with respect to the observer over the 10-year lightcurve baseline).}
    \label{fig:nom_lc}
\end{figure*}

To associate a nucleus brightness with each timestamp, we must account for the nucleus spin state and photometric properties which vary from comet to comet. In the following we describe our procedure for generating magnitudes applied to what we refer to as the \textit{nominal case}, the properties of which were chosen to represent a `typical' comet nucleus. The values chosen for the parameters of the nominal case are those of 67P, with the exception of rotation period. This we took as the mean JFC rotation period from the compilation of periods by \cite{knight_physical_2023}, $P_i=14.8978$ h, rather than the true value of $\sim$12h. This was simply to rule out the possibility of confusing real periodic signals identified in the lightcurves with alias signals due to the 24-hour day in later analysis. The values for all the initial parameters are given in Table \ref{tab:nomparams}, and those defining the nominal case are indicated in bold. 

We generated a value for the reduced magnitude H$_{0}$(1,1,$\alpha$) as a function of solar phase angle $\alpha$ at each light-time corrected timestamp using Equation \ref{eq:h0}. 

    \begin{equation}
        H_{0}(1,1,\alpha) = H(1,1,0)+\beta\alpha
        \label{eq:h0}
    \end{equation}

The absolute magnitude $H(1,1,0)$ is defined as the theoretical magnitude an object would have at heliocentric distance ($r_h$) and geocentric distance ($\Delta$) 1 au and phase angle $\alpha = 0^\circ$, where the object brightness scales linearly with phase angle by phase function $\beta$. We denote the reduced magnitude with subscript $0$ to indicate that the effects of rotation have not yet been incorporated. We used the linear phase function parameters for 67P determined by \cite{lowry_nucleus_2012} H$_{R,0}$(1,1,0) = $15.43$ 
and $\beta_L$\footnote{subscript \textit{L} to differentiate between model 
linear phase function, $\beta_L$ and initial ecliptic latitude, $\beta_i$.}$=0.059$ mag/deg.
 The LSST $grizy$ filter bandpasses are similar to those of Pan-STARRS1 (PS1), therefore we convert all magnitudes in this work to the PS1 filter system. We assumed a nucleus colour $(B-V) = 0.87$, the mean colour for JFC nuclei determined by \cite{jewitt_color_2015}, and used the conversions described in \cite{tonry_pan-starrs1_2012} to derive an absolute magnitude of $H_{r,0}(1,1,0) = 15.64$ in the PS1 $r$-band.
 
\begin{deluxetable*}{rll}
\tabletypesize{\scriptsize}
\tablecaption{Values of physical properties used to generate synthetic nucleus lightcurves. \label{tab:nomparams}}
\tablehead{
\colhead{Parameter} & \colhead{Value(s)} &\colhead{Notes}}
    \startdata
    Fixed$^*$: & \\
    \hline
    $\beta_L$ (mag/$^\circ$) & 0.059 & Linear phase function coefficient \citep{lowry_nucleus_2012} \\
     $u - r$ & 1.95 & PS1 (assuming constant spectral slope  $S^\prime=11.6 \pm 0.1 \% /1000\AA$)\\
     $g - r$ & 0.60$^a$  & $^a$Converted to PS1 from average JFC nucleus (B-V) \citep{jewitt_color_2015}\\ 
     $r - i$ & 0.35$^a$  &\\
    $r - z$ & 0.36$^a$ &\\ 
     $r - y$ & 0.41$^a$ &\\
     \hline
     Varied$^{**}$: & \\
    \hline
    Shape model & \textbf{67P/C-G} & $a/b = 1.58;\, b/c=1.15;\, a/c=1.82$\\
    & 9P/Tempel 1 & $a/b=1.29;\, b/c=1.11;\, a/c=1.42$\\
    & 103P/Hartley 2 & $a/b=3.12;\, b/c=1.13;\, a/c=3.54$\\
    & Arrokoth & $a/b=2.39;\, b/c=2.18;\, a/c=5.23$\\
    Pole obliquity $I$ ($^\circ$) & \textbf{52} & 67P pole orientation \citep{preusker_shape_2015}: $\lambda_i=78^\circ, \beta_i=41^\circ$ \\
    & $\sim$0 & Corresponds to pole direction $\lambda_i=310^\circ, \beta_i=85^\circ$\\
    & $\sim$90 & Corresponds to pole direction $\lambda_i=80^\circ, \beta_i=5^\circ$\\
    & $\sim$180 & Corresponds to pole direction $\lambda_i=130^\circ, \beta_i=-85^\circ$\\
    Period (h) & \textbf{14.8978} & \\
    & 23.9344696 & \\
    & 2.80 & \\
    & 11.10 & \\
    & 41.30 & \\
    Radius $r_n$ (km) & \textbf{2.1}, 1.0, 10.0 & \\
    Activity onset$^\dagger$ (au) & \textbf{3.0}, 0.0, 4.3 &  \\
    \hline
    \enddata
    
\tablecomments{Parameter values shown in boldtype are those which were used to generate the lightcurve for the nominal case. $^*$Parameters for which the values were kept fixed across all artificial lightcurves. $^{**}$Alternative model parameters which are varied for testing in Section \ref{params}. $^\dagger$Heliocentric distance $r_h$ below which it is assumed the comet is active. For case 0 au, the comet is inactive throughout the entire orbit i.e. points at all values of $r_h$ are included in the lightcurve dataset.}
\end{deluxetable*}

To these initial static reduced magnitudes we added the shape-driven effects of nucleus rotation.
 For the nominal case, we synthesised lightcurves using 67P shape model SHAP8, produced by spectrophotoclinometry of images taken in situ of the entire comet nucleus acquired with the Optical, Spectrocopic and Infrared Remote Imaging System (OSIRIS) instrument onboard {\em Rosetta} \citep{gaskell_characterizing_2008, jorda_global_2016}. 
 Various resolutions of this model are available - to minimise computation time, we used the lowest resolution shape model with 3000 facets. 
 
 We assigned to the shape model a constant rotation period $P_i$, and a fixed pole orientation $(\lambda_i, \beta_i$) - values are given in Table \ref{tab:nomparams}. We used the Earth-comet and Sun-comet vectors \citep{giorgini_jpls_1996} at each timestamp along with the pole orientation to determine the viewing geometry of the nucleus at each epoch. The light scattering properties of the nucleus surface were modelled using a Lommel-Seeliger (LS) function\footnote{The flux contribution from each model facet is $\frac{\mu\mu_0}{\mu+\mu_0}$, where $\mu_0$ and $\mu$ are the cosines of the angles of incidence and reflection respectively.}. Starting with an arbitrary initial rotational phase, we determined the total LS flux at each timestamp. We accounted for self-shadowing due to the non-convex shape of 67P by including the flux contributions of only facets that were illuminated, visible to the observer and unobscured by any other facets. Converting the total flux at each timestamp to a relative magnitude centred on zero gave the brightness shift to be added to the static reduced magnitudes to produce the rotational lightcurve. These reduced magnitudes were converted to apparent magnitudes by accounting for the comet's heliocentric and geocentric distances.

  To associate an apparent magnitude uncertainty with each lightcurve point, we returned to the OIF output. This includes a {\em fiveSigmaDepth} value, providing an estimate of the limiting magnitude in the specified filter at which an object would be detected with a signal-to-noise ratio (SNR) of 5. At every timestamp we compared this limiting magnitude to the corresponding apparent magnitude, first converting from PS1 $r$-band into the filter associated with that pointing. 
  We assume that the nucleus photometric properties are constant across all wavelengths. For the $grizy$ filters we converted the average JFC nucleus colours from \cite{jewitt_color_2015} to PS1 $g-r$, $r-i$, $r-z$ and $r-y$ colours using linear combinations of the polynomial colour conversions described in \cite{tonry_pan-starrs1_2012}. As PS1 has no direct counterpart to the LSST $u$-band, to estimate the colour in this band we took the average JFC ($V-R) =0.51$ from the compilation by \cite{knight_physical_2023}, noting that this is consistent with the value given in \cite{jewitt_color_2015} but based on a greater sample of 49 objects. We determined the spectral slope $S^\prime$ corresponding to this colour following the method given in \cite{luu_cometary_1990}, yielding $S^\prime=11.6 \pm 0.1 \% /1000 $\AA. Assuming a constant spectral slope from the centre of the $u$ passband to the centre of the $r$ passband, we use this value for $S^\prime$ to determine a value for $u-r$. This value is shown with the other colour indices in Table \ref{tab:nomparams}.
  
  By approximating magnitude uncertainties as 1/SNR, we extrapolated magnitude error values from the 5-$\sigma$ limiting magnitude of  each pointing. If the uncertainty of a given point was larger than 0.2, it was assumed that the object was too faint at that instance to be detected in an LSST exposure and so it was discarded from further analysis. Note that we converted the synthetic apparent magnitudes to the filter corresponding to their timestamp solely to associate a representative uncertainty value with each lightcurve point. The final lightcurve is comprised of only $r$-band magnitudes and incorporates all points with sufficient signal-to-noise to be detected by LSST over the entire ten-year time frame. 
  
  Thus far we have assumed at all timestamps that the brightness is due entirely to reflected light from the nucleus surface, which is not the case for most JFCs. For targeted observations of comet nuclei, the standard strategy to minimise contamination from activity-related brightening is to plan and undertake observations only when the comet is further than 3 au from the Sun where temperatures are insufficient for the 
  significant
  sublimation of water ice. This strategy is not without flaw, and the PSF of all targeted observations should be carefully checked to ensure that the comet's radial profile is stellar-like. To simulate this strategy, for the nominal case lightcurve we implement a simple heliocentric distance cut-off, discarding lightcurve points where the comet is at $r_h<3.0$ au. 
  
  This gives a final nucleus lightcurve as might be acquired by LSST, shown in the right panel of Figure \ref{fig:nom_lc}. In the final dataset, the median number of lightcurve points per night is 2 for the 167 individual epochs in which the nucleus is both visible and detected. The comet spans a range of $183^\circ$ to $298^\circ$ in ecliptic longitude, $-4^\circ$ to $2^\circ$ in latitude and a phase angle range $0.2^\circ$ to $16.7^\circ$.

\section{Analysis of the nominal lightcurve}
\label{analysis}

\subsection{Simple phase function fit and period search}
\label{simple}

With the synthetic sparse-in-time lightcurves for the nominal case in hand, our aim was to establish whether the known input parameters could be reproduced by standard methods of lightcurve analysis. One of the most exciting prospects for nucleus studies with LSST is the unprecedented observational coverage across entire JFC orbits. In particular, measuring a nucleus phase function has historically required years of targeted data collection at as large as possible a range of solar phase angles. However, with LSST it will be possible to gather calibrated, temporally-sparse observations spanning a large $\alpha$ range for many objects. 

\begin{figure}[t]
    \centering
    \includegraphics[width=0.5\textwidth]{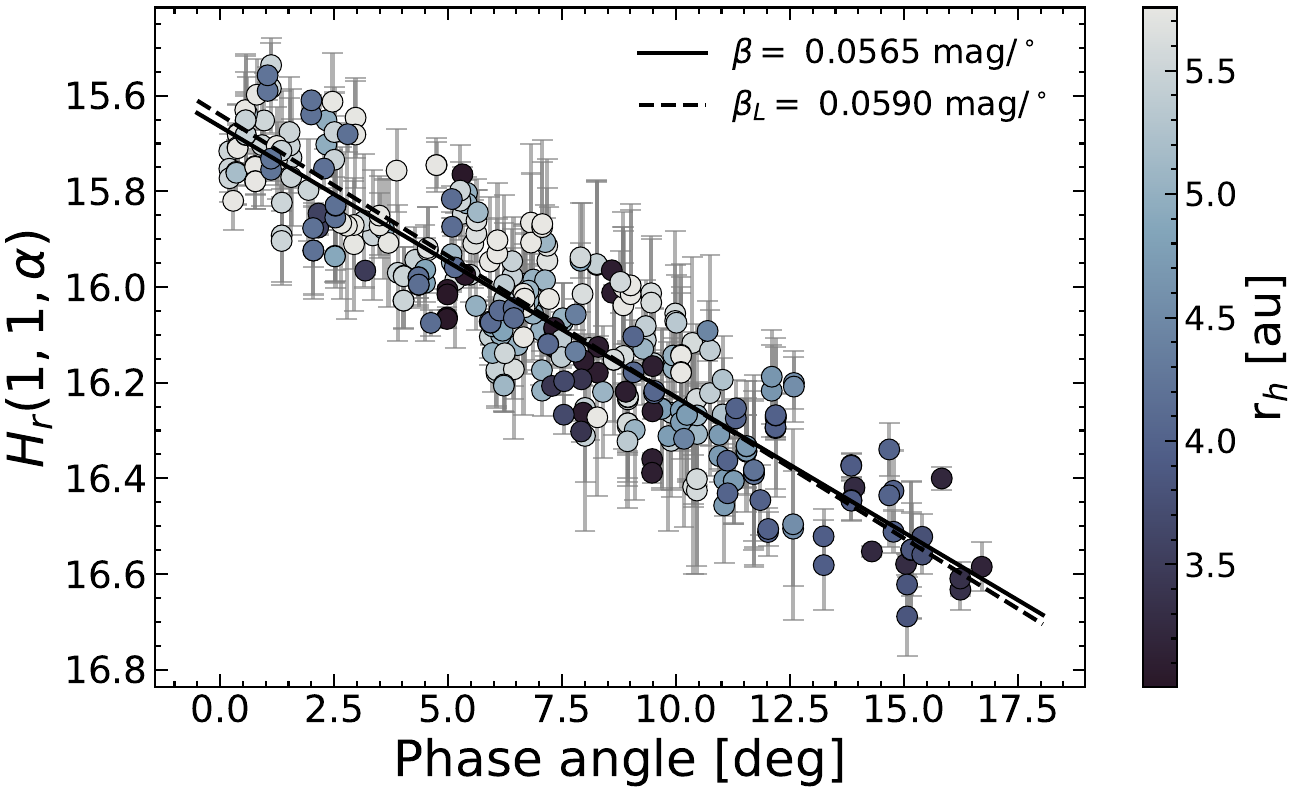}
    \caption{Reduced magnitude as a function of phase angle $\alpha$ for the nominal lightcurve. Solid line gives best fit to lightcurve points from linear regression with a slope value $\beta=(0.0565\pm0.0014)$ mag/$^\circ$. Dashed line is phase function used as input (slope $\beta_L$) to generate lightcurves. Scatter in points is due solely to lightcurve amplitude, no photometric noise incorporated at this stage and error bars are not used to weight the phase function fit.}
    \label{fig:nom_phase_nc}
\end{figure}

\begin{figure}[t]
    \centering
    \includegraphics[width=0.45\textwidth]{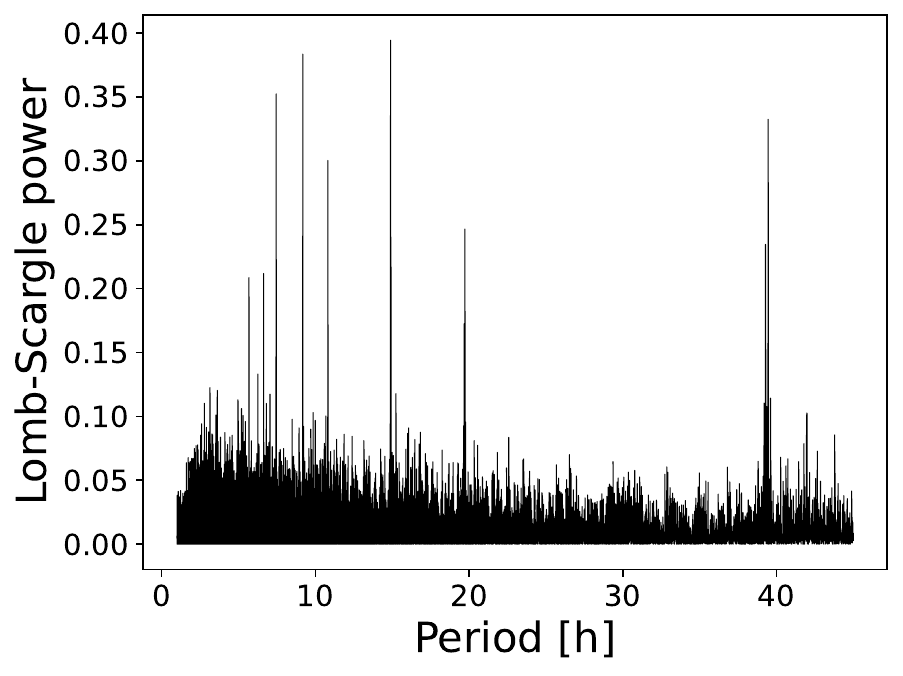}
    \caption{Lomb-Scargle periodogram for single-peaked rotation period search on nominal lightcurve. The highest peak corresponds to a most-likely synodic rotation period $P=14.898088$ h.}
    \label{fig:nomperiod1}
\end{figure}

The phase angle variation covered by the nominal lightcurve is shown in Figure \ref{fig:nom_phase_nc}, spanning $0.2^\circ-16.7^\circ$ at $r_h>3$ au. Note that there is no noise in this dataset (this is considered in Section \ref{mc}) - scatter in the points is due solely to nucleus rotation. We fit a phase function by linear regression to the reduced magnitudes, yielding a phase function coefficient $\beta=0.0565\pm0.0014$ mag/$^\circ$, slightly shallower than the input value $\beta_L = 0.059$ mag/$^\circ$. We corrected the points for this phase function by Equation \ref{eq:h0}, and searched for the best rotation period by performing a Lomb-Scargle (LS) period search \citep{lomb_least-squares_1976, scargle_studies_1982}. 
LS is typically used to determine synodic rotation periods for observational datasets in which the object geometry is unchanging. Performing a period search over the long time-base lightcurve yields an estimate for the `average' synodic period.
We tested a single-term Fourier fit over a frequency space corresponding rotation periods between $\sim$1 h and 45 h. The resulting periodogram is shown in Figure \ref{fig:nomperiod1}.  The highest periodogram peak corresponds to a rotation period $P=14.8981$ h. This is close in value to the model input sidereal period $P_i=14.8978$ h, and phasing the lightcurves to the most likely Lomb-Scargle period results in a similar profile to that shown in the right panel of Figure \ref{fig:nom_lc}. With the \textit{a priori} knowledge of the model inputs, this is consistent with the input sidereal period since the difference between synodic and sidereal periods is typically small for such observations \citep{harris_lightcurves_1984}. For observational datasets, due to the large range of nucleus observing geometries that occur over the long temporal baseline, it is challenging to select the best period from the possible LS peaks by visual inspection of the lightcurve morphology, particularly regarding the choice of a single or double-peaked lightcurve.

\subsection{Convex lightcurve inversion}
\label{clinom}

To test the benefits of the availability of a large range of observing geometries, we next assessed the synthetic lightcurves using convex lightcurve inversion (CLI). The rotation period, shape and aspect of an object directly impact the characteristics of its lightcurve measured by the ground-based observer, and thus as the observing geometry of the object changes across its orbit, so too does the appearance of its lightcurve (as can be seen in Figure \ref{fig:nom_lc}). CLI \citep{kaasalainen_optimization_2001-1, kaasalainen_optimization_2001} uses a Levenberg-Marquardt minimisation algorithm to iteratively fit the shape and spin state that most closely match the input lightcurves. This method is most effective when a wide range of object viewing geometries is covered by the input lightcurves, a diversity that is provided naturally by intermittent detections with survey telescopes. Convex inversion of solely sparse-in-time lightcurves is capable of producing extremely reliable estimates of asteroid spin states and good approximations of their overall shapes \citep{kaasalainen_physical_2004}. The primary requirement is that each data point across the entire baseline is calibrated to the same absolute brightness scale. The sparse points can then be combined and treated as a single lightcurve with a long temporal baseline.

For this work, we used the implementation of the \texttt{convexinv} software package for sparse lightcurves \citep[see e.g.][]{hanus_study_2011}.
The lightcurve inversion software parameterizes an object by its rotation period ($P$), pole orientation (ecliptic longitude $\lambda$ and latitude $\beta$), light scattering properties and shape attributes. In the optimization process the shape is represented by the expansion of a spherical harmonic series. The parameters to be optimised are the coefficients of the series with degree $l$ and order $m$. To strike a balance between reconstructing a detailed shape and avoiding over-fitting a model to a relatively small number of lightcurve points, we used $l=m=6$, giving 49 shape parameters to be optimised. 
The light scattering function is described by an empirical three-coefficient linear-exponential phase function \citep{kaasalainen_optimization_2001}. Following the method of \cite{durech_reconstruction_2023} the exponential coefficients of the phase function were held fixed, fitting solely the linear coefficient of the phase function. 

\begin{figure*}
    \centering
    \includegraphics[width=0.85\textwidth,trim=5cm 5cm 5cm 6cm]{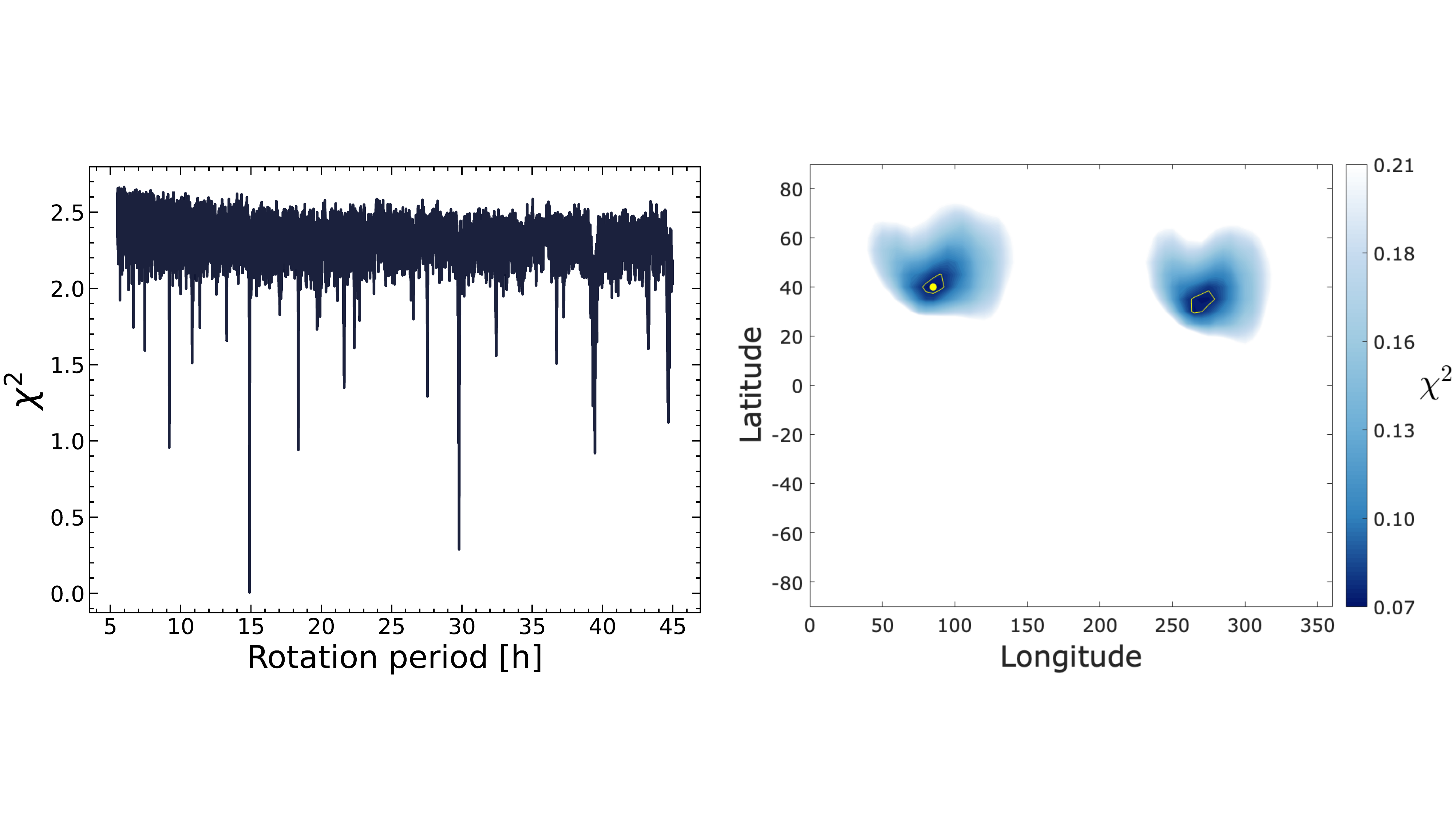}

        \caption{Left: periodogram for period search using the nominal lightcurve. The $\chi^2$ minimum is consistent with $P_i$ and the next strongest peaks occur at $P_i$ multiples and sub-multiples. Other peaks are due to sampling aliases. Right: distribution of rotation pole solutions. Darker regions indicate a lower value of $\chi^2$ and therefore a better fit to the input lightcurves. Yellow dot depicts the ecliptic coordinates corresponding to the pole orientation with best fit to the lightcurves, yellow circles show regions within which $\chi^2$ values are within 10\% of this minimum value.}
        \label{fig:nomCLI}
\end{figure*}

We searched initially for the best-fitting sidereal rotation period. The \texttt{periodsearch} program within \texttt{convexinv} steps through a series of user-defined trial periods and iteratively fits a shape at each period from six starting pole orientations. The quality of the fit to the lightcurves at each trial period is quantified by a $\chi^2$ value. The best period value is taken as the trial period corresponding to the minimum value of $\chi^2$. Figure \ref{fig:nomCLI} shows the resulting periodogram from trial periods in the range $5-45$ h. The best period was found to be consistent with $P_i$, with a value of $P = 14.8978\pm0.0001$ h, where we express the uncertainty as the period values within 10 per cent of the $\chi^2$ minimum \citep{rozek_physical_2022}. The uncertainties found by this method are generally very small as a result of the small period step size \citep[described by e.g.][]{kaasalainen_physical_2004}.

We next searched for the optimal orientation of the object's rotation pole, with the starting value of $P$ set to the above value. At each trial pole orientation on a $5\times5$ deg$^2$ grid of ecliptic coordinates, the shape was optimised over 50 iterations. The $\chi^2$ evaluation of the fit quality for each possible pole location is shown in the right panel of Figure \ref{fig:nomCLI}. There are two regions of equally-likely pole orientations, separated by approximately 180 deg in ecliptic longitude, as is expected for an object on a low-inclination orbit by the ambiguity theorem \citep{kaasalainen_inverse_2006}. We defined unique pole solutions as those with a $\chi^2$ value at least 10 per cent lower than the surrounding solutions, and which were separated from other minima by at least 30 deg. This resulted in two possible pole orientations corresponding to the regions of darkest blue shown in Figure \ref{fig:nomCLI}. Solution 1 at $\lambda=84^\circ$, $\beta=40^\circ$ is consistent with the input coordinates used for the pole orientation, and Solution 2 at $\lambda=274^\circ$, $\beta=35^\circ$ gives the mirror solution. These values are permitted to vary as part of the final shape optimisation procedure described below, but testing showed that the pole orientation at which the final model converges is close to the coordinates of the minima presented here. Therefore, to associate a formal uncertainty with these pole orientations, we consider the uncertainty region as that with $\chi^2$ values within ten per cent of the $\chi^2$ minimum value. This yields final pole solutions of $\lambda=(84\pm6)^\circ$, $\beta=(40\pm5)^\circ$ for Solution 1, and $\lambda=(274\pm7)^\circ$, $\beta=(35\pm5)^\circ$ for Solution 2.

Solution 1 can be directly compared to the results of other published convex inversions of 67P lightcurves: the first acquired using dense-in-time photometry by \cite{lowry_nucleus_2012} who identified a pole orientation ($\lambda,\beta)=(78^\circ\pm10^\circ,58^\circ\pm10^\circ$);
and with the addition of lightcurves from the approach phase of the {\em Rosetta} spacecraft, \cite{mottola_rotation_2014} obtained a most likely pole orientation ($\lambda,\beta)=(65^\circ\pm15^\circ,59^\circ\pm15^\circ$).
These solutions are consistent with one another within their specified uncertainty ranges, and with the true orientation of 67P's rotation pole.

\begin{figure}[t]

    \centering
    \includegraphics[width=0.48\textwidth]{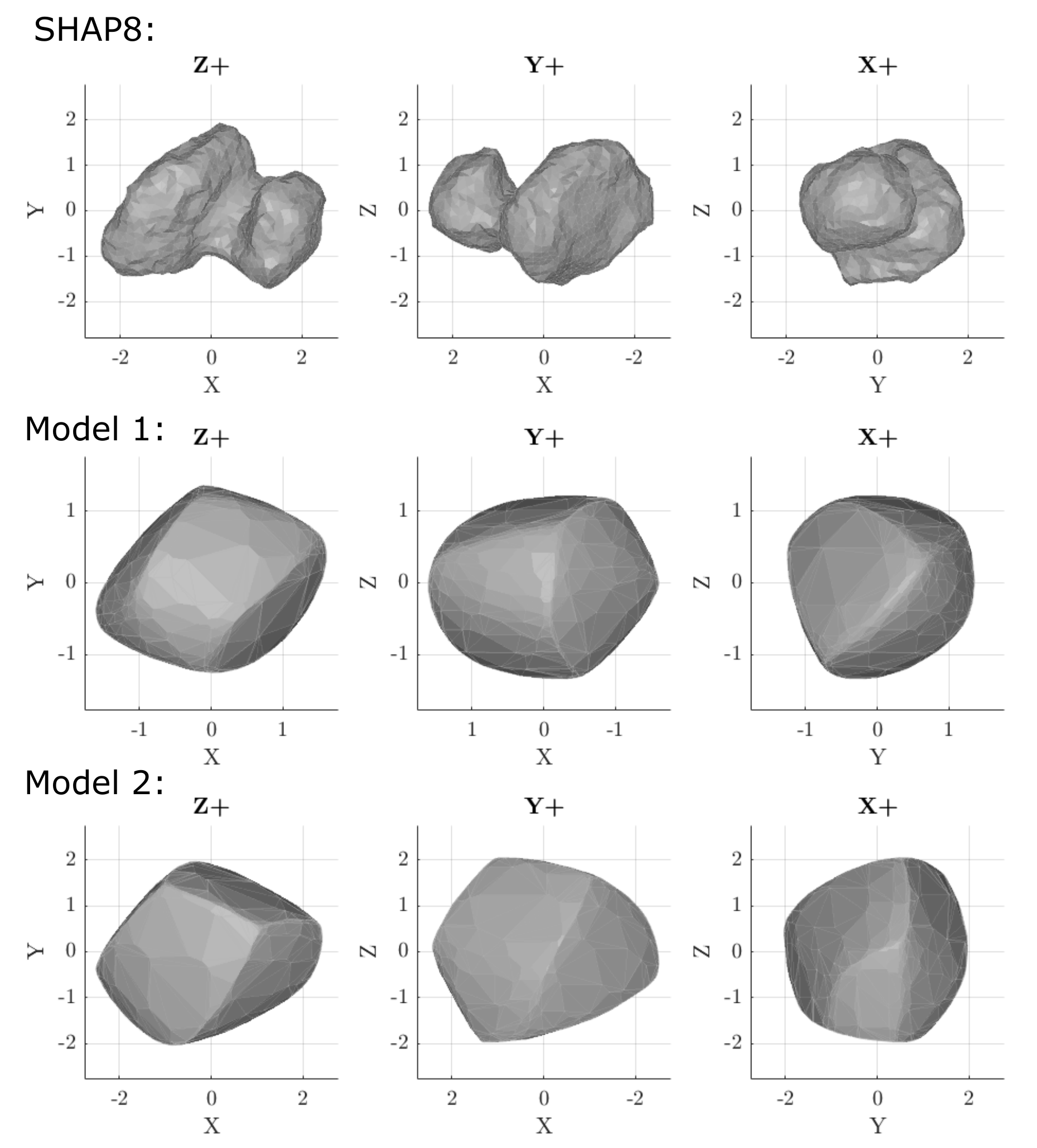}

        \caption{A comparison of input 67P shape model (top) with the final convex models corresponding to pole Solution 1 (centre) and the mirror at Solution 2 (bottom). Three orthogonal views are given for each model, with the viewing angle oriented along the axis indicated in bold above the model. The rotation axis is aligned with the Z-axis.}
        \label{fig:nomshapes}
\end{figure}

For each of these pole orientations, we developed a convex shape model. We used 50 iterations of the Levenberg-Marquardt loop to fit each model, unless the $\chi^2$ values had not satisfactorily converged within this number of iterations. In most cases it was found that iterating beyond 50 when the $\chi^2$ values had already converged (varied by $<10^{-3}$ with each iteration) made minimal difference to the axis ratios and fit to the lightcurves of the final shape models, and only served to make the models `blockier' and less physically realistic in appearance, likely due to the small number of lightcurve data points  (typically 1-2) to fit to at each geometry. The two final shape models are shown in Figure \ref{fig:nomshapes}. The axis ratios of Model 1 are $a/b=1.15$ and $b/c=1.15$, and for Model 2 $a/b=1.13$ and $b/c=1.16$. 

The axis ratios of the convex models are substantially less elongated than those of the input 67P model (see Table \ref{tab:nomparams}). This is to be expected for an object as strongly non-convex as 67P - the lightcurve inversion procedure cannot recreate surface concavities and instead tends to mask over them, which has the effect of artificially lengthening the rotation axis of the convex model in this case. We test the impact of different nucleus shapes later in Section \ref{shape}. It is therefore more illustrative to compare these shapes to the previously published convex models of 67P obtained by \cite{lowry_nucleus_2012} and \cite{mottola_rotation_2014}. The axial ratios based on the inertia tensors of these models are $a/b=1.19, b/c=1.25$ and $a/b=1.14, b/c=1.22$ respectively. 
The sparse-in-time model axis ratios derived in this work are remarkably similar to the previously published models. This is extremely promising for characterising a greater number of JFCs with LSST photometry: the diversity of observing geometries covered over the survey's 10-year period of operation is capable of providing shape and pole information that is equivalent to years of targeted monitoring.


\begin{figure*}
    \centering
    \includegraphics[width=0.7\textwidth]{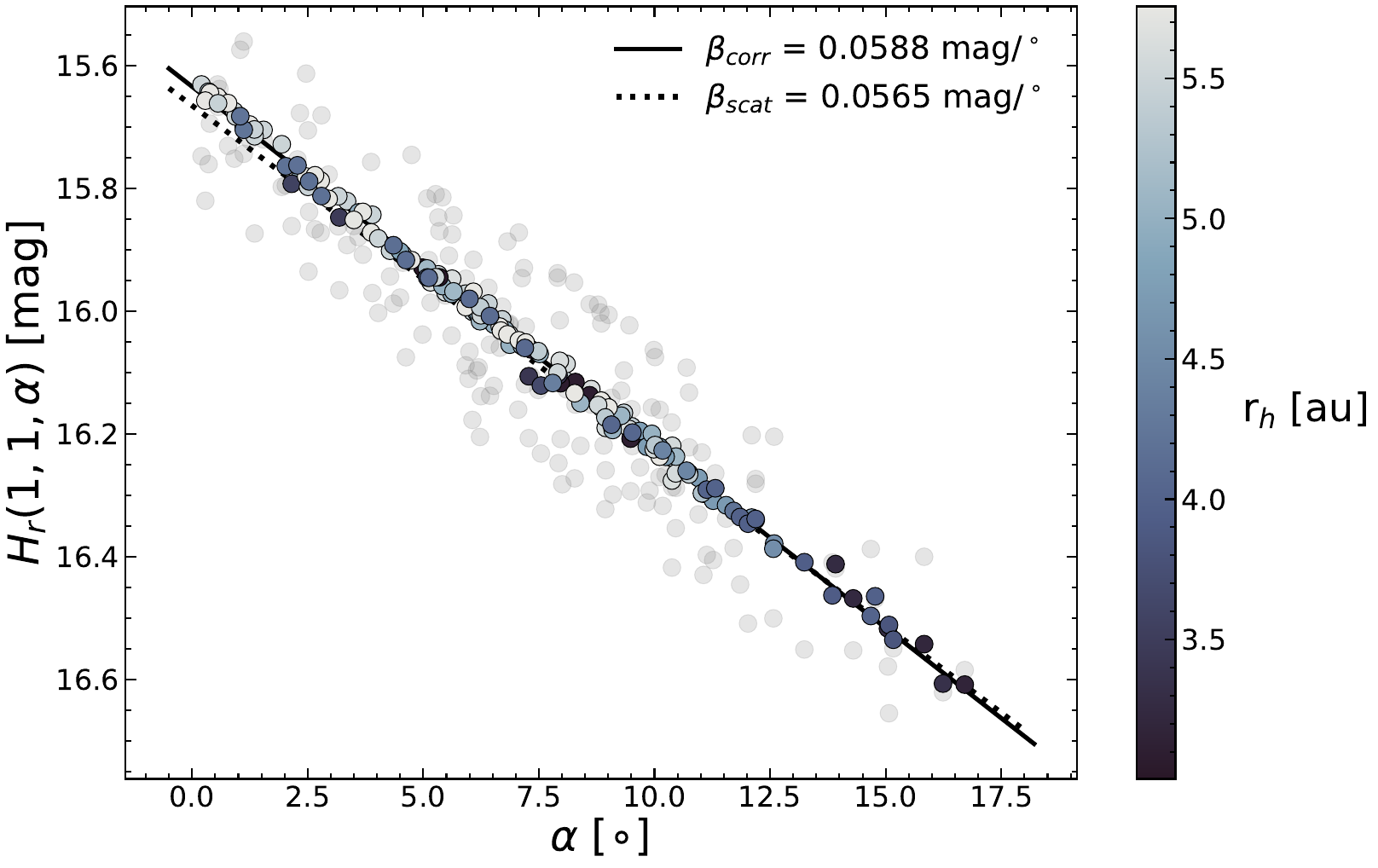}
    \caption{Nucleus phase function for the nominal lightcurves. Blue-coloured points show the synthetic magnitudes corrected for the effects of rotation by the Solution 1 shape model, coloured according to heliocentric distance r$_h$. The translucent grey points depict the lightcurve points before the rotational correction. The solid line shows the best linear fit to the corrected points (slope value $\beta_{corr} = 0.0588\pm0.0002$ mag/deg, intercept $H(1,1,0) = 15.634\pm0.002$), and the dashed line the fit to the original, scattered points (slope value $\beta_{scat} = 0.0565\pm0.0014$ mag/deg, intercept $H(1,1,0) = 15.665\pm0.012$).}
    \label{fig:phasefunc}
\end{figure*}

While the convex models can provide valuable insights into the object's global shape properties, these insights are somewhat limited when the original shape is intrinsically non-convex, such as that of 67P. One can convince oneself that Model 1 is a gift-wrapped rendering of the shape of the bilobed nucleus, close to a faithful reconstruction of its convex hull. In contrast, since the model was optimised to fit each input lightcurve point, we can use the model to correct the lightcurve points for the effects of brightness variation due to the rotation of the irregular shape \citep[analogous to the \textit{reference} phase function of][]{kaasalainen_optimization_2001}. Following the method of \cite{donaldson_characterizing_2023}, all lightcurve points within a single rotation were binned and shifted by the mean correction. The shifted magnitudes were plotted as a function of phase angle $\alpha$, shown in Figure \ref{fig:phasefunc}. While the phase function of the rotationally-scattered points is slightly shallower than the input phase function $\beta_L$, following the rotational-correction with the convex model $\beta_L$ is exactly recovered.

\section{Testing a range of cometary properties}
\label{params}

We have thus far tested LSST-like photometry for a single combination of possible comet physical properties. In reality, the properties of the well-studied JFCs span a wide range of values. We selected five parameters to vary that we considered likely to have the largest impact on the simulated lightcurves: rotation period; pole obliquity; nucleus shape; nucleus size; and heliocentric distance cutoff (below which it is assumed that bare nucleus observations are not possible due to the dominance of sublimation-driven activity in photometry apertures). This section aims to test whether the intrinsic nucleus properties affect the likelihood that the `correct' results, i.e. the input parameters, are reproduced by lightcurve analysis. 

For each test case, we regenerate a nucleus lightcurve by changing only one model parameter at a time. 
The parameter variant tests are described below, and the final lightcurve analysis results for all cases can be found in Table \ref{tab:paramres}. The starting rotation of the model is kept as the same arbitrary time $T_0$ as was used for the nominal case. In every test case, we found that the pole search returned two equally-likely unique solutions for the pole orientation, corresponding to the best solution and its mirror as described in Section \ref{clinom}. For brevity, in the following we report only the pole coordinates and shape parameters of the solution which corresponds to the `correct' orientation of the rotation pole, and we discuss the implications of the pole solutions further in Section \ref{discpole}.

\begin{deluxetable*}{l|ccccccccccc}[t]
\tabletypesize{\scriptsize}
\tablecaption{Resulting values for parameter fits when model inputs were varied. \label{tab:paramres}}
\tablehead{\colhead{Varied parameters} & \colhead{Pole Lon. ($\lambda$)} & \colhead{Pole Lat. ($\beta$)} & \colhead{$\Delta m_{max}$} & \colhead{$\Delta$Ampl.$^1$} & \colhead{$\beta_{scat}$} & \colhead{$\Delta\beta_{scat}$$^2$} & \colhead{$\beta_{corr}$} & \colhead{$\Delta\beta_{corr}$$^2$} & \colhead{$a/b$} & \colhead{$b/c$} & \colhead{$a/c$}\\
\colhead{} & \colhead{($^\circ$)} & \colhead{($^\circ$)} & \colhead{(mag)} & \colhead{(mag)} & \colhead{(mag/$^\circ$)} & \colhead{(\%)} & \colhead{(mag/$^\circ$)} & \colhead{(\%)} & \colhead{} & \colhead{} & \colhead{}}
\startdata
    $r_h>3.0$ au & $84\pm6$ & $40\pm5$ & 0.43 & 0.13 & $0.0565\pm0.0014$ & 4.2 & $0.0588\pm0.0002$ & 0.3 & 1.15 & 1.15 & 1.32 \\
    $r_h>0.0$ au & $81\pm6$ & $38\pm5$ & 0.51 & 0.21 & $0.0587\pm0.0008$ & 0.5 & $0.0596\pm0.0002$ & 1.0 & 1.06 & 1.21 & 1.33 \\
    $r_h>4.3$ au & $90\pm20$ & $37\pm13$ & 0.40 & 0.12 & $0.0544\pm0.0020$ & 7.8 & $0.0584\pm0.0004$ & 1.0 & 1.15 & 1.20 & 1.38 \\
    \hline
    r$_n \sim$ 1km & $85\pm5$ & $34\pm8$ & 0.41 & 0.13 & $0.0574\pm0.0024$ & 2.7 & $0.0595\pm0.0007$ & 0.8 & 1.11 & 1.22 & 1.36  \\
    r$_n \sim$ 10km & $82\pm8$ & $39\pm5$ & 0.42 & 0.13 & $0.0567\pm0.0012$ & 3.9 & $0.0588\pm0.0003$ & 0.3 & 1.12 & 1.17 & 1.31 \\
    \hline
    $P_i = 2.8$h & $82\pm5$ & $41\pm5$ & 0.41 & 0.14 & $0.0581\pm0.0014$ & 1.5 & $0.0595\pm0.0002$ & 0.8& 1.16 & 1.18 & 1.38 \\
    $P_i = 11.1$h & $81\pm4$ & $41\pm4$ & 0.38 & 0.11 & $0.0593\pm0.0015$ & 0.5 & $0.0586\pm0.0003$ & 0.7 & 1.16 & 1.14 & 1.32 \\
    $P_i = 41.3$h & $83^{+12}_{-3}$ & $42\pm3$ & 0.40 & 0.11 & $0.0556\pm0.0014$ & 5.8 & $0.0586\pm0.0003$ & 0.7 & 1.17 & 1.18 & 1.39 \\
    $P_i = 23.9344696$h & $81_{-6}^{+75}$ & $41^{+4}_{-25}$ & 0.37 & 0.13 & $0.0564\pm0.0007$ & 4.4 & $0.0583\pm0.0003$ & 1.2 & 1.22 & 1.14 & 1.39 \\
    \hline
    Shape: 9P & $78^{+32}_{-8}$ & $37\pm3$ & 0.24 & 0.12 & $0.0585\pm0.0008$ & 0.8 & $0.0579\pm0.0004$ & 1.9 & 1.10 & 1.05 & 1.14 \\
    Shape: 103P & $79\pm2$ & $42\pm2$ & 0.86 & 0.62 & $0.0542\pm0.0018$ & 8.1 & $0.0575\pm0.0007$ & 2.5 & 1.70 & 1.36 & 2.31 \\	
    Shape: Arrokoth & $80_{-10}^{+25}$ & $33\pm8$ & 0.29 & 0.24 &$0.0594\pm0.0004$ & 0.7 & $0.0588\pm0.0003$ & 0.3 & 1.15 & 1.12 & 1.29 \\
    \hline
    Obliquity: $I = 0^\circ$ & $319^{+30}_{-40}$ & $86^{+4}_{-26}$ & 0.40 & 0.08 &$0.0528\pm0.0016$ & 10.5 & $0.0592\pm0.0003$ & 0.3 & 1.26 & 1.60 & 2.01 \\
    Obliquity: $I = 180^\circ$ & $183^{+82}_{-22}$ & $-85^{+20}_{-5}$ & 0.42 & 0.09 & $0.0564\pm0.0017$ & 4.4 & $0.0601\pm0.0003$ & 1.9 & 1.19 & 1.18 & 1.40 \\	
    Obliquity: $I = 90^\circ$ & $79\pm5$ & $5\pm5$ & 0.33 & 0.31 & $0.0573\pm0.0010$ & 2.9 & $0.0588\pm0.0003$ & 0.3 & 1.11 & 1.14 & 1.26 \\
\enddata

\tablecomments{$^1$ Range of model lightcurve amplitudes i.e. $\Delta m_{max}-\Delta m_{min}$. $^2$Absolute percentage difference between measured phase function and input phase function $\beta_L$. The large uncertainties in pole orientation for cases $P_i=23.93..$ h, $I=0^\circ$ and $I=180^\circ$ correspond to error ellipses with equivalent radii $24^\circ$, $23^\circ$ and $25 ^\circ$ respectively.}
\end{deluxetable*}

\newpage
\subsection{Activity distance}

To examine how cometary activity impacts the nucleus lightcurves, we implemented a simple threshold to dictate the minimum $r_h$ value at which lightcurve points could be included in the final dataset. For the nominal case, this threshold was set at 3 au. We tested two additional cutoffs: one at 4.3 au, the heliocentric distance of 67P below which which its activity could be detected by ground-based facilities \citep{snodgrass_beginning_2013}; and one at 0 au i.e. the comet was inactive for the duration of its orbit. The latter case could be representative of a devolatised JFC at the end of its dynamical lifetime \citep{2004jewittcradle}. 

Generating synthetic lightcurves for $r_h>0$ au and $r_h>4.3$ au as outlined in Section \ref{sim_mags}, the number of data points remaining in the final lightcurves was 423 and 221 respectively. Including points across the entire orbit meant a maximum phase angle $\alpha_{max}=26^\circ$ was covered by the lightcurves, and the initial phase function fit to these scattered lightcurves was a close match to the input value of $\beta_L$. Discarding all points below 4.3 au reduced the maximum observed phase angle to $\alpha_{max}=12.6^\circ$, yielding a slope value of $\beta_{scat} = (0.054\pm0.002)$ mag/deg, 7.8 per cent shallower than the input value of $\beta_L$. 

For both cases, the period search component of the convex lightcurve inversion identified $P_i$ as the period corresponding to the global $\chi^2$ minimum. The pole searches successfully identified coordinates for the pole positions that were consistent with the simulation inputs. For the case $r_h>4.3$ au, it was found that the uncertainty in the pole coordinates was larger than for the nominal case - examination of the distribution of $\chi^2$ values over the entire ecliptic sphere showed a large range of coordinates had a $\chi^2$ value within $1.1\times$ the minimum. This is likely a result of the reduced number of data points in the lightcurve covering a narrower range of nucleus aspects - in the dataset, both the nominal case and $r_h>0$ au cover an aspect angle range of $\sim46^\circ$, while imposing an $r_h>4.3$ au cutoff covers only $30^\circ$ in aspect. 

The axis ratios of both final shape models are similar to those found for the nominal case, implying that the inversion procedure robustly identifies the optimal shape when lightcurve points are both added and removed. This indicates that a sufficient number of observing geometries are covered by the lightcurves in the $r_h>4.3$ au test case, that the final shape and spin solution is not noticeably improved with the addition of lightcurves closer to the Sun. The rotationally-corrected the phase functions $\beta_{corr}$ for both cases deviated from $\beta_L$ by only 1 per cent, and the uncertainty in the phase function fit was reduced.

\subsection{Size}

We next considered the effect of nucleus size on our ability to extract nucleus properties from the LSST photometry. The effective nucleus radius can be expressed in terms of its geometric albedo $p_r$ and mean absolute magnitude $\bar{H}_r$ by Equation \ref{eq:radius}:

\begin{equation}
    r_n = k/\sqrt{p_r}\times 10^{0.2(m_\odot-\bar{H_r})}
    \label{eq:radius}
\end{equation}

The constant $m_\odot$ refers to the brightness of the Sun in the filter of measurement, and $k=1.496\times10^8$ km per au. 
For an effective radius of approximately $1$ km, we use a value of $\bar{H}_r=17.1$ mag, and for $10$ km $\bar{H}_r=12.2$ mag. 
These radius values were chosen to order-of-magnitude approximate the extremes of the size range of the currently-known JFCs \citep[see Table 1A of][]{knight_physical_2023}. 

The lower limit of 1 km was imposed by our method of lightcurve generation - we initially trialled 0.5 km as the small size case, but found that this resulted in the disposal of 97 per cent of the initial 613 timestamps due to the low signal-to-noise of the lightcurve points at large $r_h$. This left fewer observed points than model parameters to fit in the lightcurve inversion process, and so we opted to scale the small radius up to 1 km. The final synthetic lightcurves consisted of 91 and 515 lightcurve points for the $r_n\sim1$ km and $r_n\sim10$ km models respectively.

For both cases, the initial phase function fit to the scattered lightcurve points was 3-4 per cent shallower than the initial $\beta_L$ value. The period and pole searches correctly identified the input period and pole orientation for both size models - this is particularly noteworthy for the small value of $r_n$ with only 91 data points total. The shape models were very similar. In both cases, using the convex model to correct the phase curve for rotational scatter allowed the input phase function $\beta_L$ to be precisely recovered, within the uncertainties on the slope.

\subsection{Rotation period}

Varying the rotation period allows us to examine the effects of spin rate on the quality of the recovered nucleus properties. To sample a representative range of JFC rotation periods, we generated lightcurves using nucleus models rotating with the minimum ($P_i=2.8$ h), maximum ($P_i=41.3$ h) and median ($P_i=11.1$ h) rotation periods for all known JFCs \citep[from the compilation in][Table 1A]{knight_physical_2023}. We also tested a nucleus with a spin rate identical to Earth's sidereal period ($P_i=23.9344696$ h) as an extreme case of sampling only a fraction of the object's lightcurve profile. 

As was the case for the previous tests, the lightcurve inversion successfully identified the input period value as the most likely rotation period for each trial lightcurve, and the pole searches returned the input pole orientation. 
This was true also for challenging case of $P_i\sim24$ h, although the distribution of $\chi^2$ values for all pole coordinates tested (Figure \ref{fig:chi24}) suggests four regions of similar probability as a result of the ambiguous shape of the lightcurve. We have accounted for these additional possibilities in the uncertainties quoted for the pole coordinates in Table \ref{tab:paramres}.

The initial phase function estimate for the $P=41.3$h lightcurves deviated significantly from $\beta_L$ by 5.8 per cent and the pole search revealed a relatively large uncertainty in polar longitude. The latter finding can be attributed to the sparse observations sampling comparatively less of the rotational phase of the slow rotator, leading to increased uncertainty in the estimate of the body meridian.  

\vskip 0.2in
\subsection{Shape}
\label{shape}

The choice of 67P as the typified JFC nucleus with which to simulate lightcurves allowed us to probe whether the non-convex shape model still allowed us to constrain physical parameters such as period and pole. 
The meagre sample of five JFCs with resolved nucleus imaging (9P, 19P, 67P, 81P and 103P) exhibit a wide diversity of shapes. As such, 67P was a natural choice of shape model as a bilobed object that is not particularly elongated. 

To examine the impact of nucleus shapes on the lightcurve analysis outcomes, we trialled three additional shape models: JFCs 9P/Tempel 1 and 103P/Hartley 2; and trans-Neptunian object (486958) Arrokoth. Comets 9P and 103P represent the two extremes of the known JFC shapes, with 9P being irregularly-shaped but primarily rounded \citep{farnham_plate_2013}, while 103P is extremely elongated and dogbone-like in shape \citep{farnham_103p}. The shape model of Arrokoth is based on images returned by the NASA New Horizons mission \citep{stern_initial_2019}, which revealed a bilobed object with an extremely large $a/c$ axis ratio. To date there have been no in situ detections of a JFC with such a shape.

For 9P and Arrokoth, the initial phase function fit to the scattered lightcurve points already matched $\beta_L$ well within the slope uncertainties. 
In both cases, $P_i$ was successfully identified as the best period and the input pole orientation was recovered by the lightcurve inversion, however the uncertainties associated with the pole longitude are sizeable. This is likely a result of the small range of lightcurve amplitudes covered: for 9P, the lightcurve $\Delta m$ ranged from $0.12-0.24$, and for Arrokoth $0.05-0.29$. 
The convex model for 9P was slightly less elongated than the true shape (axis ratios given in Table \ref{tab:nomparams}) but is visually similar to the true shape (see Figure \ref{fig:shapes}). 
In contrast, the convex model for Arrokoth was an extremely poor match to the input shape model. The lightcurve inversion favoured a close to spherical shape to fit the input lightcurves (see Figure \ref{fig:shapes}). 
This is likely due to the extremely low $\Delta m$ values of the lightcurves. 
The phase function fits following correction by the convex model were comparable to the initial scattered fit within their uncertainties. 

The initial phase function fit to the 103P lightcurves resulted in a shallow $\beta_{scat} = 0.0542\pm0.0018$ mag/$^\circ$, more than 8 per cent shallower than $\beta_L$. The lightcurve inversion successfully determined the period and constrained the pole orientation to within a very small region of uncertainty. For the shape optimisation, we found it was necessary to iterate 400 times before the $\chi^2$ fits to the lightcurves began to converge. The inversion identified an elongated shape for 103P (Figure \ref{fig:shapes}), though the axis ratios were substantially less elongated than the true shape by nature of the convex model artificially elongating the rotation axis. The difference between $\beta_{corr}$ and $\beta_L$ was reduced to 2.5 per cent following rotational correction.



\subsection{Pole}

Spin axis orientation is known for only a small number of JFCs and is typically constrained by tracking evolving coma morphology. For low-activity comets, pole orientation is coupled with shape information, and thus simultaneous shape and pole optimisation by lightcurve inversion, such as demonstrated here, is one of the few ways of estimating this parameter for such objects. 

For the above tests we used the real pole orientation of 67P \citep{preusker_shape_2015} which corresponds to an obliquity\footnote{Angle subtended by the orbital momentum vector (perpendicular to orbit plane) and the vector in the direction of the spin axis.} $I\sim52^\circ$. For a given orbit, obliquity 
has a large impact on the measured lightcurve, as the region of the nucleus surface reflecting sunlight towards the observer changes as the polar aspect varies across the orbit. To test this effect, we synthesised lightcurves for nuclei in stable rotation states with obliquities of $I\sim0^\circ$, $I\sim180^\circ$ and $I\sim90^\circ$, i.e. the polar direction aligned with the orbital normal vector in both the positive and negative directions, and pole perpendicular to the orbit normal vector. The ecliptic longitude and latitude values used to generate these lightcurves for 67P's orbit are given in Table \ref{tab:nomparams}.

\begin{figure}
\centering
\noindent
\begin{tabular}{c}
        \includegraphics[trim={8cm 0cm 6cm 1cm},clip,width=0.4\textwidth]{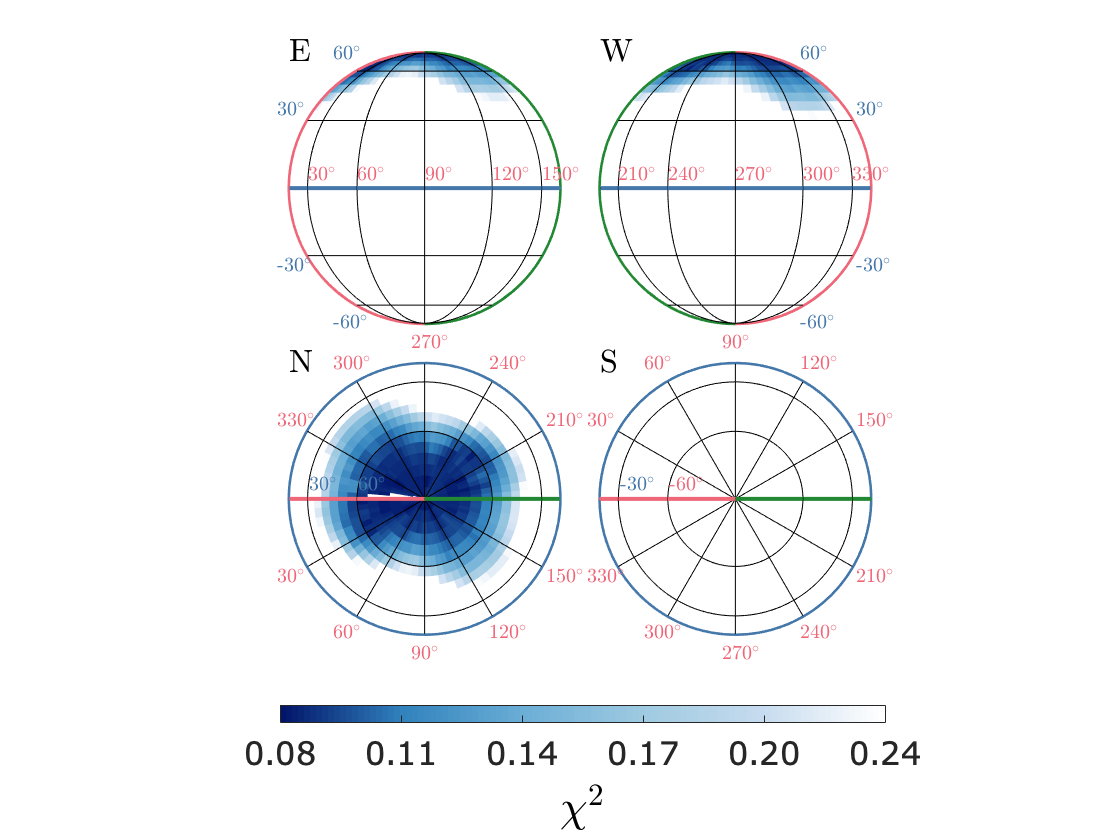} \\
        (a) \\
       \includegraphics[trim={8cm 0cm 6cm 1cm},clip,width=0.4\textwidth]{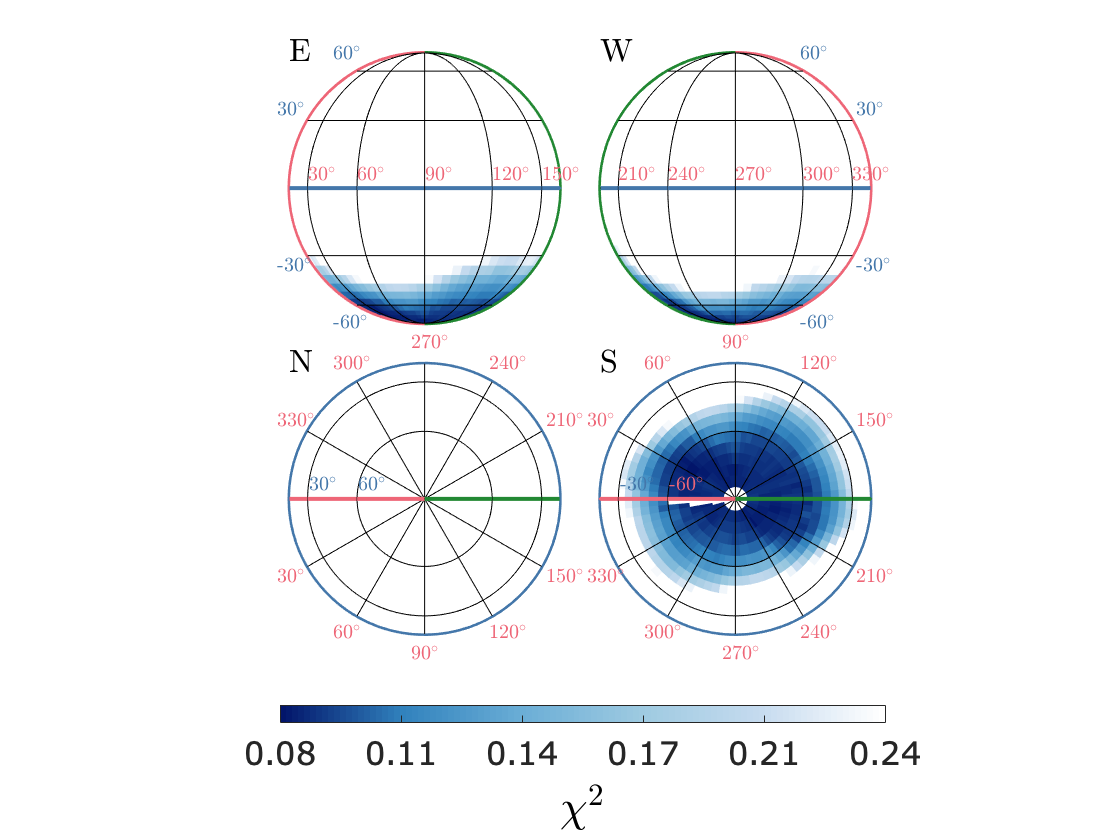} \\
        (b) \\
\end{tabular}
\caption{Results from 5$\times$5 deg pole search on synthetic lightcurves with pole obliquity (a) $I=0^\circ$ and (b) $I=180^\circ$ projected onto the ecliptic sphere.}
    \label{fig:0dpn}
\end{figure}

For all three pole orientations, the $\beta_{scat}$ fit was notably shallower than $\beta_L$. The best period was again consistent with $P_i$ in every case. For each test the pole search was able to converge to two unique solutions. 
At the maximum obliquity angle of $I=90^\circ$, the pole orientation is tightly constrained and the convex shape properties are comparable with those from the previous test cases. 
For the $I\sim0^\circ$ and $I\sim180^\circ$ tests, the unique pole solutions were close to the ecliptic poles where longitude values are more closely spaced, shown in Figure \ref{fig:0dpn}. 
The regions of uncertainty on the pole span $20-25^\circ$ degrees in ecliptic latitude, centred on the poles. 
While there is seemingly more ambiguity in the pole coordinates when considering the uncertainty regions, there is very little physical difference between different longitude values at ecliptic latitudes close to $\pm90^\circ$. 

The axis ratios of the best-fit shape for the $I=0^\circ$ test deviate most from the nominal case than any other varied parameter. The same cannot be said for the opposing obliquity at $I=180$, but visual inspection of the best-fit model reveals a physically improbable shape with more than 100$^\circ$ offset between the rotation pole and the shortest axis of inertia of the shape's equivalent-volume ellipsoid.
These effects are likely due to the limited variation in object geometry caused by the low obliquity. Across these datasets the pole aspect varies by approximately $4^\circ$, resulting in only small portions of their surface ever being illuminated to the observer. It is unlikely that any physically-meaningful shape information can therefore be extracted from lightcurves of low-inclination objects with low obliquity. The shape models were still effective for correcting the lightcurves for rotation, and in both cases brought the $\beta_{corr}$ fit much closer to the true value.


\begin{figure*}
\centering
\noindent
\begin{tabular}{c c}
        \includegraphics[width=3.2in]{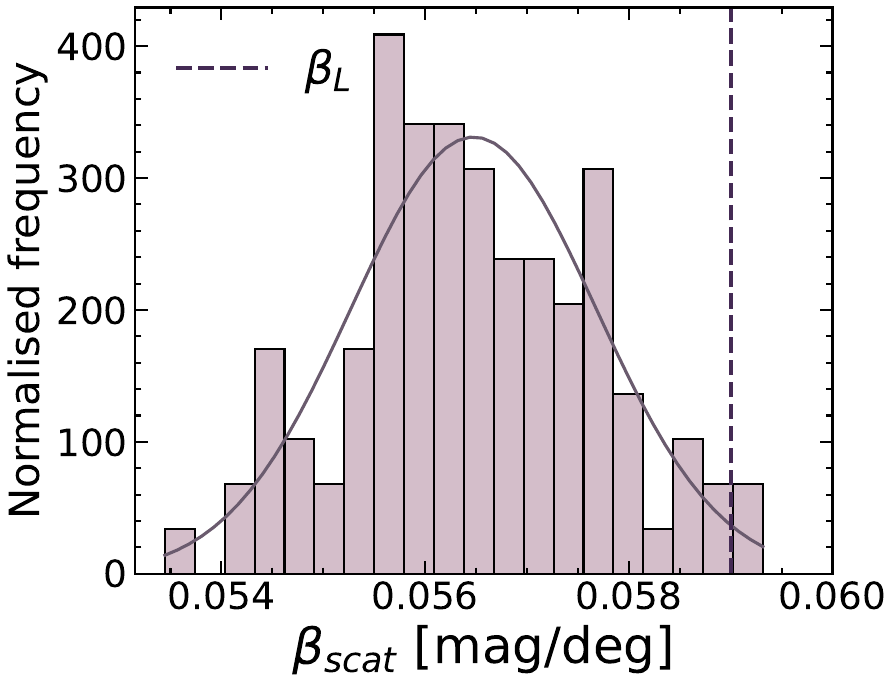} &
            \includegraphics[width=3.05in]{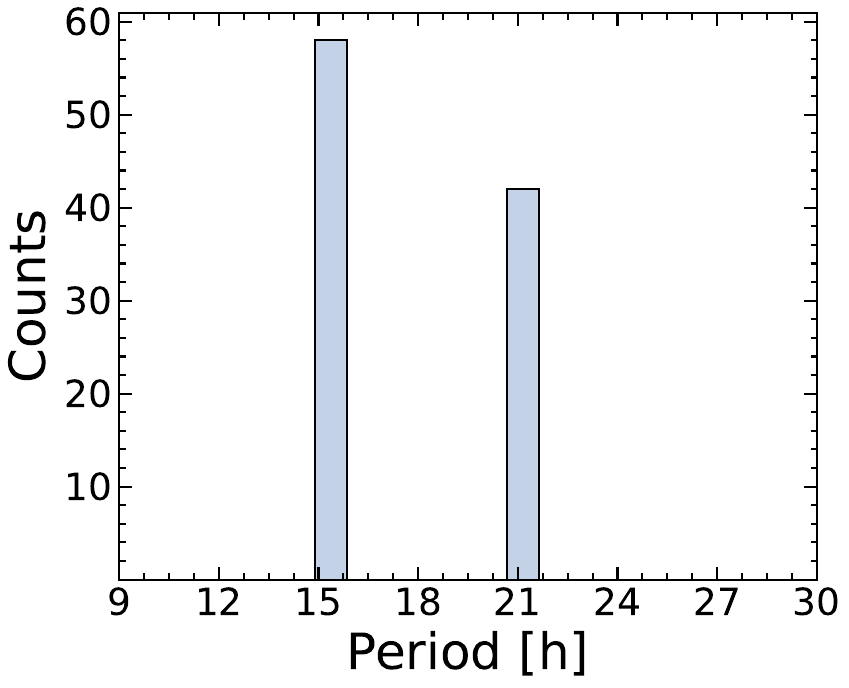} \\
            (a) & (b) \\
\end{tabular}
\caption{Histograms from the Monte Carlo (a) phase function and (b) period analysis for the nominal case. The true phase function slope $\beta_L$ is given by the dashed vertical line in the left panel. The period values shown are twice the most-likely period value identified by the Lomb-Scargle analysis.}
    \label{fig:bp_nom}
\end{figure*}

\vskip 0.2in

\section{Incorporating photometric noise}
\label{mc}

\subsection{Sparse lightcurves}
We demonstrated above that convex lightcurve inversion reproduces rotation periods and pole orientations extremely effectively
when the lightcurves precisely describe the input model parameters - i.e. the lightcurve points include no photon noise. The small sizes of JFCs mean they are faint (20--24 mag) at the large heliocentric distances typically required to observe them with minimal activity, thus real nucleus brightness measurements will include significant photon noise. In the following we attempt to evaluate the effect of photon noise on our ability to extract nucleus parameters from LSST JFC observations using a Monte Carlo (MC) approach. Noisy versions of the nominal lightcurve are generated by replacing each lightcurve point with a randomly-generated magnitude clone taken from a normal distribution with mean and standard deviation respectively equal to the apparent magnitude and uncertainty value determined in Section \ref{sim_mags}. We repeat this process creating $n=100$ lightcurves to sample a large range of the noisy parameter space.

We first tested the simple case of fitting a phase function and rotation period to the sparse lightcurves, as in Section \ref{simple}, for $n=100$ lightcurves. For each lightcurve we transformed the randomised apparent magnitudes back into reduced magnitudes $H_r(1,1,\alpha)$, and used weighted linear regression to fit the phase function $\beta_{scat}$ as a function of $\alpha$. The resulting distribution is shown in the left panel of Figure \ref{fig:bp_nom}. A normal distribution fit to this histogram yields a mean value $\bar\beta_{MC} = 0.0566\pm0.0012$ mag/$^\circ$. 

Each randomised lightcurve was converted to absolute magnitudes by its respective $\beta_{scat}$ value, and we performed a Lomb-Scargle period search as described in Section \ref{simple}. The distribution of best synodic period values is shown in the right panel of Figure \ref{fig:bp_nom}. The period most often identified by the Lomb-Scargle algorithm (60 per cent of tests) was consistent with $P_i$, and a period of $\sim21$ h was preferred for approximately 40 per cent of the lightcurves. We found that across all 100 lightcurves, the range of possible values for the best period identified by the algorithm was very small - within each histogram bin, the period values were similar to within 2 decimal places. This was true even for period searches with a large number of frequency samples per peak, and is likely a result of the sparse rotational sampling of the data over an extensive temporal baseline.

\begin{figure}[h]
    \centering
    \includegraphics[width=0.45\textwidth]{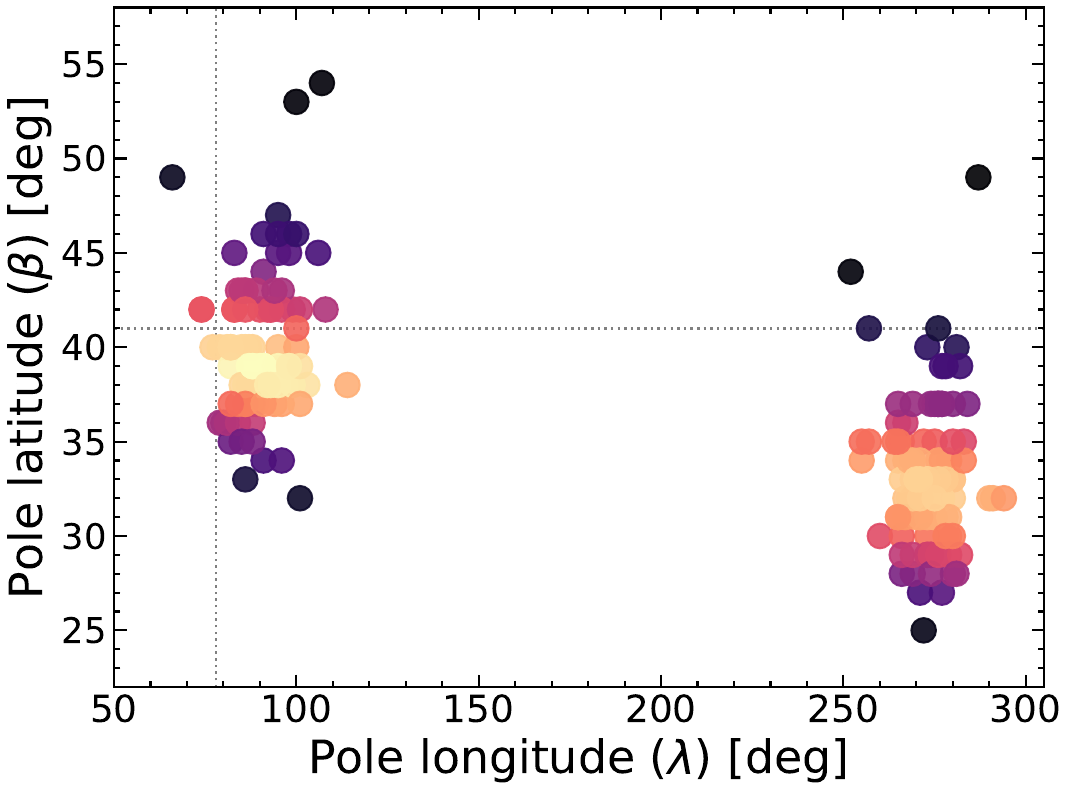}
    \caption{Pole orientations corresponding to the top two unique pole solutions from lightcurve inversion of 100 sparse-in-time lightcurves. Colour corresponds to density of points where lighter regions have a larger number of points. Dashed lines mark the input pole coordinates ($\lambda_i$, $\beta_i$).}
    \label{fig:MCpoles}
\end{figure}

\begin{figure}
    \centering
    \includegraphics[width=0.46\textwidth]{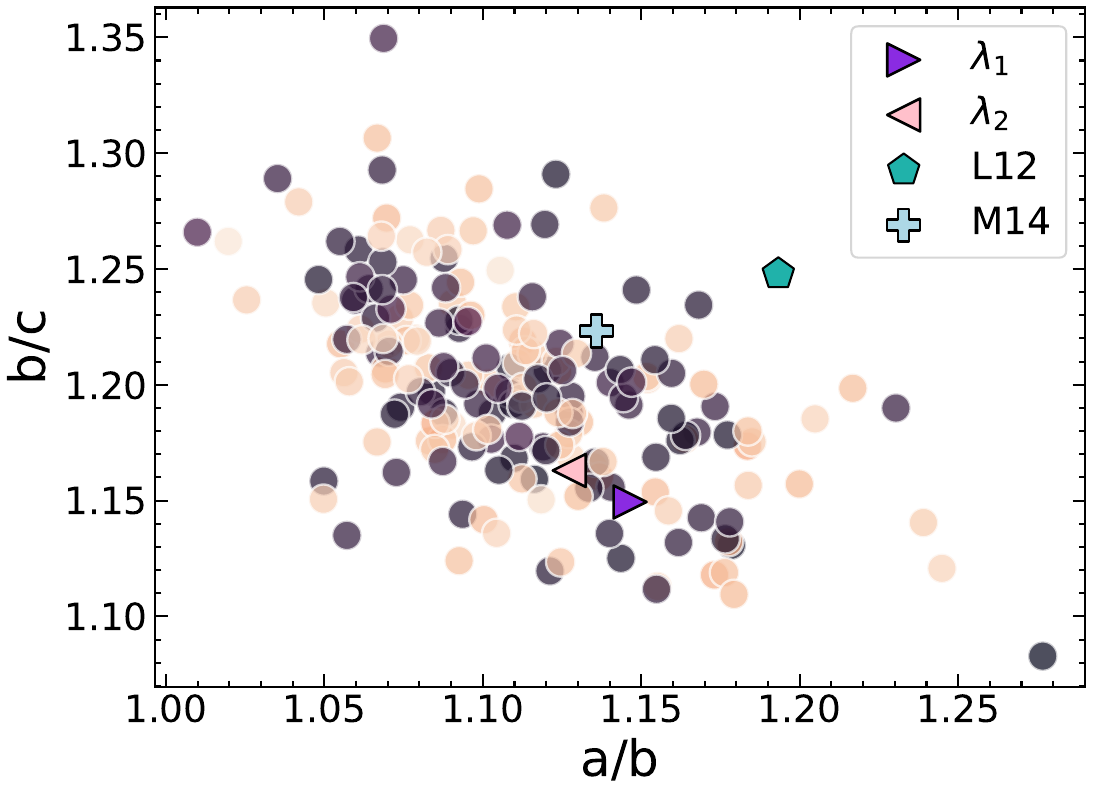}
    \caption{Distribution of principal axis ratios $a/b$ and $b/c$ for the top two shape models from 100 MC randomised lightcurve trials. Colour scale depicts pole longitude: darker coloured points correspond to pole orientations closest to the simulation input pole, lighter colours are shapes at the mirror pole. Also shown are axis ratios for the noiseless models developed in Section \ref{analysis}, and previously published models by Lowry et al. (\citeyear[L12]{lowry_nucleus_2012}) and Mottola et al. (\citeyear[M14]{mottola_rotation_2014}).}
    \label{fig:paMC}
\end{figure}

Applying convex inversion to the random lightcurves, we repeated the procedure outlined in Section \ref{clinom} with 100 newly-generated lightcurve clones. We found that for every lightcurve, a period was identified that was close to $P_i$, with the values ranging from $14.8976-14.9018$ h. For 36 of the 100 lightcurves, a period multiple in the range $29.7953-29.8033$ h was identified as equally likely, highlighting the challenge of correctly identifying the period from noisy data. We performed the pole search for each random lightcurve with a starting period value of $P_i$, and allowed this value to be optimised along with the pole over 300 Levenberg-Marquardt iterations. The distribution of the top two unique pole solutions for all lightcurves is given in Figure~\ref{fig:MCpoles}. Considering only the models consistent with the input pole orientation ($\lambda_i=78^\circ,\,\beta_i=41^\circ$) i.e. not the mirror solutions, the mean pole was located at ($\lambda=91^\circ,\,\beta=40^\circ$) with interquartile ranges for the longitude and latitude values of $10.5^\circ$ and $4^\circ$ respectively. This implies that it is highly likely that after 10-years of the LSST, the rotation periods and pole orientations for a large number of JFCs will be known to a high degree of certainty. We optimised the convex shapes at the top two pole solutions for every lightcurves. Distributions of the model axes ratios are shown in Figure \ref{fig:paMC}. Most models are similar in elongation to the noiseless models, and lie within $1.05<a/b<1.25$.

For every random lightcurve, we selected the convex model corresponding to the pole direction closest to the true orientation, to correct for rotation and fit the phase function. The $\beta_{scat}$ and  $\beta_{corr}$ distributions are shown in Figure \ref{fig:all_betas}. We found that the phase function values shifted positively to be centred on $\beta_L$ following the correction, with a mean and standard deviation of $\bar\beta_{corr,s}=0.0591\pm0.0015$ mag/$^\circ$. While the correction improves the likelihood of correctly estimating the phase function, the shape model fit to the noisy data points resulted in an increased spread in the final distribution of $\beta_{corr}$ values.

\subsection{Addition of dense-in-time lightcurves}
As a final test, we repeated the MC convex inversion with the addition of dense-in-time lightcurves. Lightcurves covering large segments of the rotational phase help to constrain the rotational parameters at that geometry, and the inclusion of a single dense lightcurve has been demonstrated to measurably improve inversion outcomes for sparse asteroid photometry \citep[e.g.][]{santana-ros_testing_2015}. 

We created synthetic lightcurves across three dark nights in April 2029 when the comet would be observable with a $2$-metre class telescope in the northern hemisphere for around half of the night. In these `observations', the target is at a heliocentric distance of $3.6$ au, spanning phase angles $4.6-5.2^\circ$ with a pole aspect $\sim$$80^\circ$.
We estimated the magnitude uncertainties as 0.04 mag, and created 100 random instances of these lightcurves to be combined with the randomised LSST points. As the uncertainties in the dense lightcurves were approximately half as large as the average uncertainty in the sparse lightcurve, we weighted them to contribute twice as much as the sparse lightcurve to the fit at each stage of the optimisation process. 

\begin{figure*}
    \centering
    \includegraphics[width=0.6\textwidth]{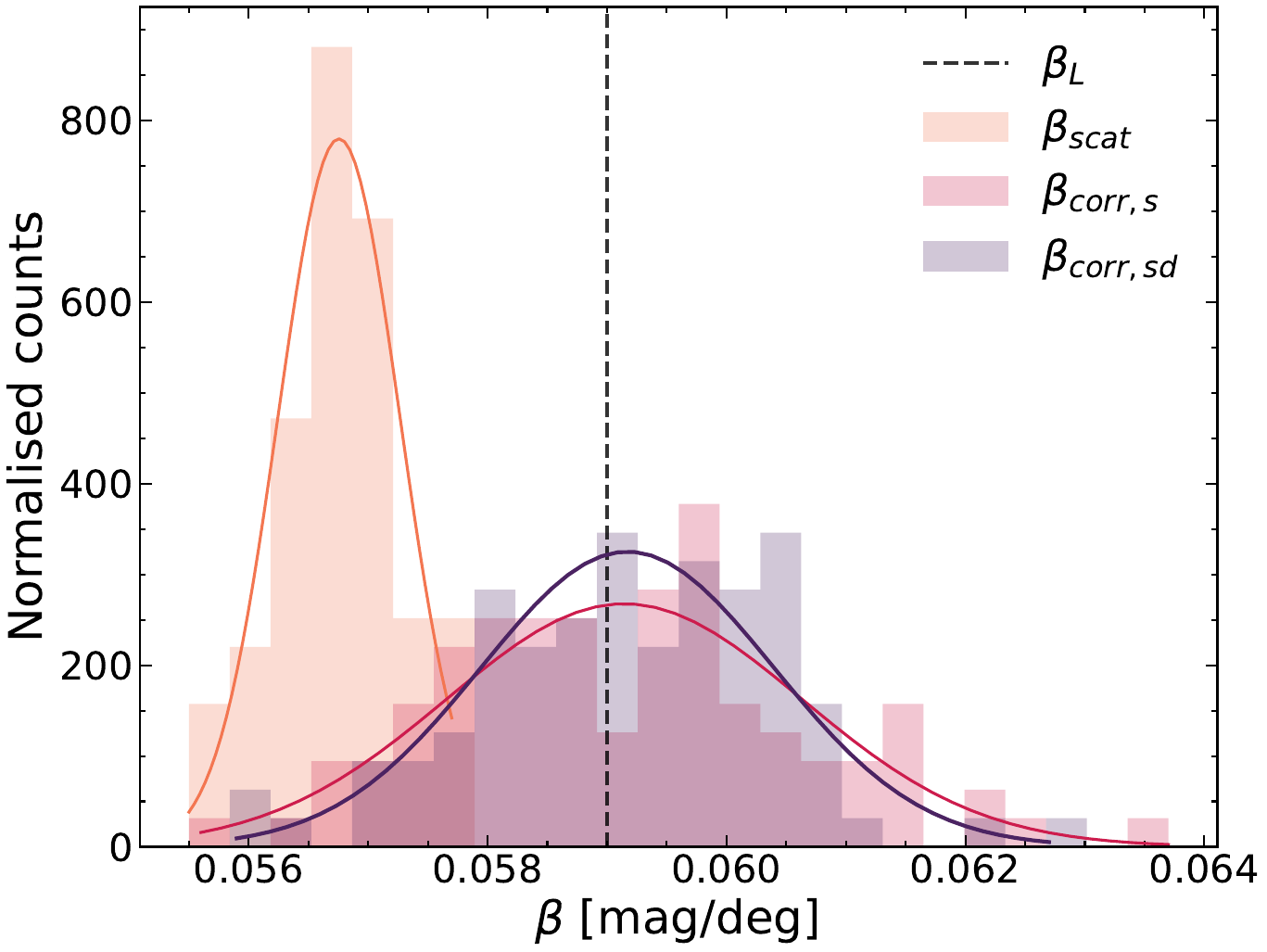}
    \caption{Distribution of phase function fits to 100 MC lightcurves for three different cases. Dashed vertical line shows the input phase function value $\beta_L$. Orange line ($\beta_{scat}$) demonstrates weighted linear regression on reduced magnitudes with lightcurve amplitude scatter. The mean of this distribution is $0.0567\pm0.0005$ mag/$^\circ$. Red line ($\beta_{corr,s}$) shows distribution of phase functions for lightcurves corrected for rotation by a convex model constructed from purely sparse lightcurves with mean $0.0591\pm0.0015$ mag/$^\circ$. Purple line (labelled $\beta_{corr,sd}$) results from the phase function fit using a convex model constructed from combination of sparse and dense lightcurves, with mean $0.0591\pm0.0012$ mag/$^\circ$.}
    \label{fig:all_betas}
\end{figure*}

\begin{figure}[t]
    \centering
    \includegraphics[width=0.45\textwidth]{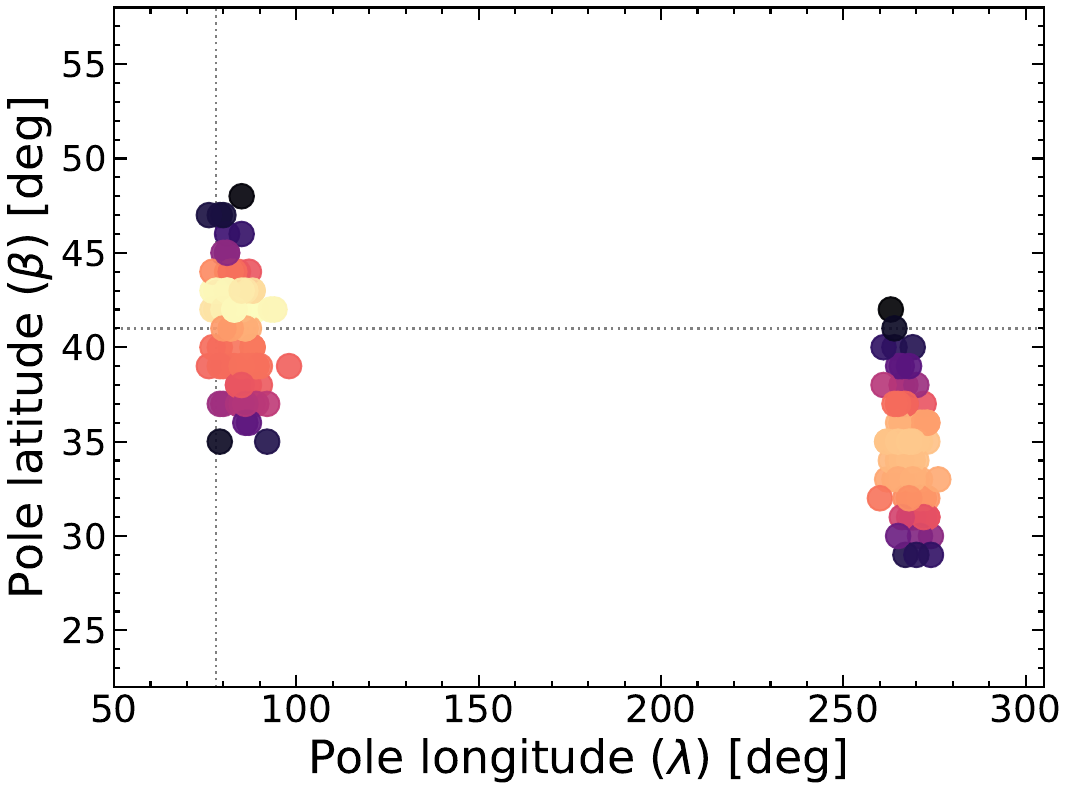}
    \caption{Same MC analysis as in Figure \ref{fig:MCpoles} but with the addition of a dense lightcurve to the sparse survey data points.}
    \label{fig:MCdense}
\end{figure}

Across all 100 of the combined dense and sparse lightcurves, the period search identified values consistent with $P_i$ as the best-fit rotation period solutions. Unlike in the case of the solely sparse lightcurves, no alias periods were returned as equally-likely, indicating that the inclusion of dense lightcurves substantially reduced the ambiguity in the period estimation from the noisy sparse datasets. The distribution of pole orientations resulting from the pole searches is shown in Figure \ref{fig:MCdense}. Considering the points closest to the true pole orientation, the mean ecliptic coordinates are $\lambda=84^\circ, \beta=41^\circ$. The distribution of possible pole orientations are demonstrably more tightly constrained when the lightcurve inversion has access to dense-in-time photometry: compared with the sparse-only lightcurves the inter-quartile range (IQR) of the model polar latitude is unchanged at $4^\circ$, while the longitude IQR is reduced to $6^\circ$ and there is less scatter across the entire set of models. The rotationally-corrected phase function for the combined sparse and dense lightcurves is shown in Figure \ref{fig:all_betas}, with a mean and standard deviation $\bar\beta_{corr,sd} = 0.0591\pm0.0012$ mag/deg. The distribution mean is unchanged from the phase function fits to the sparse-only lightcurves, but there is a reduction in the distribution standard deviation.


\section{Discussion}
\label{disc}

\subsection{Predictions for the LSST}

The broad conclusion of this work is that accurate characterisations of nucleus properties will be possible from LSST photometry. The range of nucleus variants for which the model parameters were successfully reproduced suggests that these characterisations will enhance our understanding of JFCs on an unprecedented scale. Throughout this analysis we have made several simplifying assumptions in order to generate the synthetic lightcurves. Below we contextualise our findings outside of these assumptions, and predict what can realistically be expected from the output of LSST.

\subsubsection{When will reliable results be attainable?}
\label{long}
This analysis was based on lightcurves comprised of data points acquired across the complete operational period of LSST. This prompts the question of how much data is required before we can accurately measure the different JFC rotational properties, and at which point of the survey this amount of data will be available.
The answer to this question clearly depends upon a host of factors including the survey start date, observational cadence, and orbital properties on an object-by-object basis. 
While we cannot succinctly answer this within the scope of this analysis, we tested our ability to determine the input parameters from the lightcurves obtained for our nominal test case after one and three years of survey commencement. 
This emulated the potential data availability after the first and third annual data releases.

Comet 67P has an orbital period of 6.4 years and its perihelion occurs approximately halfway through the LSST dataset in 2028. The first three years therefore cover the comet around aphelion and inbound to $\sim4.6$ au. 
The Year 1 dataset was comprised of 51 lightcurve points spanning ecliptic longitudes of $224^\circ$ to $231^\circ$ and latitudes $-1^\circ$ to $-0.5^\circ$. The lightcurve amplitudes ($\Delta m$) ranged from 0.36 to 0.39.
For the Year 3 dataset there were 120 lightcurve points spanning $224^\circ$ to $267^\circ$ in ecliptic longitude and $-3^\circ$ to $-0.5^\circ$ in latitude. Lightcurve amplitudes were in the range 0.30 to 0.39.

For both cases with noiseless lightcurves, the most-likely rotation period was found to be consistent with $P_i$. For the Year 1 test the pole solution did not converge to a unique solution, 
but for Year 3 two unique poles were identified at 
$\lambda_1 = 280^\circ\pm15^\circ, \beta_1 = 35^\circ\pm10^\circ$ and 
$\lambda_2 = 104^\circ\pm10^\circ, \beta_2 = 40^{+5}_{-15}$. 
The latter solution is close to the input pole orientation $\lambda_i=78^\circ, \beta_i=41^\circ$

We also ran an MC period and pole search on the Year 3 lightcurve.
Across 100 random lightcurves, there were an average of 2.25 equally-likely period solutions with a maximum number of 10. Every lightcurve found a period value within 0.005 h of $P_i$ as a possible solution. For 70 per cent of the lightcurves there were one or more equally-likely rotation periods in the range $5.1 - 39.5$ h. Analysing the pole solutions as in Section \ref{mc}, using an initial period of $P_i$ and disregarding the mirror solutions, the mean pole orientation was $\lambda=104^\circ$, $\beta=26^\circ$ with an IQR of $15^\circ$ for longitude and $24^\circ$ for latitude respectively.

While there is significantly more uncertainty in the Year 3 dataset for period and pole compared with the results for the `complete' dataset described in Section \ref{mc}, these findings are encouraging and point to the possibility of measuring the nucleus rotational parameters within a single orbit.
Our method therefore indicates that, assuming observations are obtained at large heliocentric distance, it is possible for the lightcurve inversion to identify the correct period and approximate pole orientation from three years of LSST observations (or $>100$ detections with sufficient SNR).


\subsubsection{Rotation period}
\label{discper}
For every nucleus parameter variant tested in this work, the input model rotation period could be accurately recovered solely from the sparse lightcurve points. 
Incorporating photometric noise led to difficulty distinguishing between the true period and period multiples, which was alleviated with the addition of dense-in-time lightcurves.
For real nucleus photometry with no prior knowledge of the object's rotation period, it may be necessary to perform lightcurve inversion for all possible periods and rule out values by visual inspection.
Nevertheless this is extremely promising for augmenting the number of JFCs with well-constrained rotation periods, without the need for extensive targeted monitoring campaigns. 

The simplifying assumption made in this analysis was that the model nucleus was in a stable spin state for the survey duration, rotating with a constant period and pole orientation.
It is well established that JFCs can experience period changes on measurable timescales \citep[e.g. ][their Table 2 and references therein]{knight_physical_2023}. 
Such spin changes are likely to occur close to perihelion where sublimative activity is maximised, and empirical models have indicated that the spin rates of smaller, more active nuclei are more strongly impacted than larger nuclei \citep{samarasinha_relating_2013, jewitt_systematics_2021}.

In the event that significant spin changes take place, employing sparse long-time baseline observations such as those simulated here may aid in constraining the extent of the variation. 
We have shown that changing nucleus geometry can be satisfactorily accounted for by lightcurve inversion, and that the rotation period can be extracted from a single orbit (Section \ref{long}).
Therefore, if a rotation period is measured by lightcurve points obtained over a time frame including a perihelion passage, and it produces a poor fit to lightcurve points, this may indicate a change in spin rate.
This approach was used to detect the Yarkovsky-O'Reefe-Radzievskii-Paddack (YORP) effect in near-Earth asteroids from dense-in-time photometry \citep{kaasalainen_acceleration_2007}. YORP induces linear-in-time variation in the spin states of these objects in contrast with the step function around perihelion that is exhibited by JFCs. 

Given that the magnitude of spin change is believed to scale inversely with nucleus size, we expect that if spin changes do take place throughout the operational period of LSST, the nuclei for which the effect is strongest will be too faint for lightcurve inversion by LSST (see discussion below).
 Most JFCs will make at most one perihelion passage over the ten-year survey duration.
Based on our findings we anticipate that accurate rotation periods will be accessible for many km-sized JFCs, provided they have been observed at large heliocentric distance for a significant portion of the survey. If the observations encompass a perihelion passage, we do not rule out the possibility of period changes being detectable by lightcurve inversion.




\subsubsection{Shape and pole}
\label{discpole}

The strength of lightcurve inversion is the simultaneous optimisation of shape and pole in addition to rotation period.
We found that in the test cases with a low obliquity angle, the pole orientation was correctly constrained to its respective ecliptic hemisphere. For every test case with an intermediate pole obliquity, the best-fit pole orientation returned the input direction as well as the $\lambda+180^\circ$ mirror solution. Where there was a large region of uncertainty around the best-fit pole coordinates, either the lightcurves exhibited low amplitude values or, in the case of the comet rotating with $P_i\sim24$ h, there was fundamental uncertainty in the lightcurve profile. 

These findings suggest that for most low-inclination JFCs, unless they have obliquities close to $0^\circ$ or $180^\circ$, lightcurve inversion of LSST photometry will result in dual equally-likely pole orientations. However, this ambiguity is only in polar longitude as the pole latitude is similar (within $\sim10^\circ$) for both the true and mirror pole orientations. As discussed by \cite{durech_asteroid_2009}, for low-inclination orbits the obliquity can be estimated directly from the polar latitude. 

A larger sample of JFCs with known obliquities will provide insights into seasonal activity variations from ground-based observations. Constraining the rotation poles for a more representative sample of the JFC population may also allow further development of models of comet evolution. \cite{samarasinha_preferred_1997} and \cite{neishtadt_evolution_2002} predicted that the rotation poles of short period comets would tend towards a point of rotational stability oriented along the direction of maximum outgassing i.e. towards the perihelion or aphelion directions.

Regarding nucleus shapes, we had variable success in reproducing the overall shape profiles of the input shape models by convex inversion. The best-fit convex models for the lightcurves generated with a 67P shape model had comparable axis ratios to the literature convex models, with the exception of the low obliquity tests which varied in aspect by only $\sim4^\circ$. We would caution against relying on shape information from sparse data points for such objects with low orbital inclinations as a result.

The lightcurve inversion favoured more rounded shapes even when the input shape model was elongated in the cases of 103P and Arrokoth. For every case the convex model axis ratio was less elongated than the true $a/b$ value. We attribute this to the convex regularisation having the effect of masking over potential concavities and artificially extending the shorter model axes.
We expect that elongated shapes like 103P and Arrokoth are likely to require substantial dense-in-time lightcurve coverage to give the lightcurve inversion adequate constraints for a more representative shape model. 
Interpretation of how well convex shape models generated by sparse photometry represent non-convex comet nuclei is beyond the scope of this work. 

\subsubsection{Activity}

With the inclusion of a dense lightcurve in Section \ref{mc} we weighted the lightcurves' contribution to the final shape and spin model according to their uncertainties, which stem from the apparent brightness of the data points. However, for real active objects with poorly-understood activity patterns, brighter objects tend to be closer to the Sun and more likely to be contaminated by activity. Treatment of uncertainties in brightness measurements of potentially active objects should always be carefully considered alongside an assessment of their activity levels.

The simplifying assumption typically applied to JFCs (and used for the majority of the tests in this work) is that they are inactive beyond 3 au from the Sun where temperatures are insufficient for water ice to sublimate. However, 
a significant number of JFCs have exhibited activity at heliocentric distances beyond this \citep[e.g.][]{mazzotta_epifani_distant_2007, mazzotta_epifani_distant_2008,kelley_persistent_2013}.
We demonstrated that only incorporating data points acquired at $r_h>4.3$ au still produced accurate results in the noiseless test case, but the drivers of activity at large distances from the Sun are poorly understood and implementing a simple distance cutoff does not guarantee data points will be free from activity contamination, for example in the case of outbursts.
Interpreting photometric observations of comet nuclei therefore always requires a careful assessment of the comets' activity levels. 

There are various challenges in assessing the activity of faint sources, but it is an important consideration as undetected activity in photometry apertures may lead to erroneous interpretation of lightcurves.
If unaccounted for, significant coma flux could suppress the rotational signature of the nucleus and yield anomalously bright lightcurve points, which may be mistaken for shape effects if incorporated into datasets for lightcurve inversion.
This scenario will be difficult to recognise solely from sparse survey data points.
A deep activity search combining frames acquired at similar geometries may reveal unresolved coma.
It is likely that the frequent detections and signal-to-noise capabilities of LSST for many JFCs at large $r_h$ will significantly enhance our understanding of activity at such distances - for example, calibrated brightness measurements across large portions of the orbit may aid in the future identification of anomalous brightening.


\subsubsection{Nucleus size limitations}
\label{discsize}

Using our method of synthetic lightcurve generation, we found it was necessary to use a minimum effective radius of 1 km. Setting the low end of the nucleus size range to 0.5 km, which more closely reflects the size of the smallest observed JFC nuclei, left only $\sim20$ data points for lightcurve analysis. This meant fewer data points than model parameters to fit by the chosen method of lightcurve inversion, and thus we opted not to test it in this work. 

If the true photometric uncertainties in the survey are larger than our simple SNR approximation, it is possible that only those JFCs with $r_n>1$ km will be detected by the survey at the large heliocentric distances typically necessary for nucleus characterisation. 
Using the \cite{bauer_debiasing_2017} debiased JFC size distribution, and accounting crudely for the fraction of inactive JFCs as 75 per cent those passing aphelion during the LSST that do not exhibit activity at large heliocentric distance \citep{kelley_persistent_2013}, we estimate that this analysis will be possible for a few hundred JFC nuclei.
For objects with $r_n\sim0.5$ km, 20 lightcurve points at a range of geometries may be sufficient to deduce a first approximation to the rotation period and pole orientation if a model with fewer fit parameters is employed. These could be refined with targeted follow-up observations by a designated observing program to characterise small nuclei discovered by LSST. 


\subsection{Phase functions}

Varying the input parameters for the nucleus model in Section \ref{params}, it was found that the phase function fit $\beta_{scat}$ to the reduced magnitudes was often not consistent with the input phase function value $\beta_L$. The average difference ($\Delta\beta_{scat}$) between the phase function fit to the scattered lightcurves and $\beta_L$ across all varied parameters was 4 per cent. This deviation can be attributed simply to rotational scatter: fitting a linear function to data with $y$-axis scatter yields greater uncertainty in the slope (and intercept) values. Even with an MC approach to sample the distribution of phase function fits to the noisy lightcurves as in Figure \ref{fig:bp_nom}, recovering the true slope value was unlikely. 

In most cases, the difference between $\beta_L$ and the measured $\beta_{scat}$ was insignificant, but for several cases the deviation was of order 10 per cent and the phase function measured by this method may lead to erroneous interpretations of the nucleus surface properties. The large deviation in the 103P slope value is likely a result of the high lightcurve amplitudes ($\Delta m$) which, when sampled by the survey simulator, resulted in a larger spread of possible magnitude values for the data points. A similar argument can be applied to the phase function fit to the $I=0^\circ$ lightcurves: such an obliquity for an object with a low-inclination orbit would result in a mostly equatorial view of the rotating nucleus. For 67P, this implies viewing the bilobed nucleus face-on, resulting in long periods at the flattened lightcurve maxima and hence more scatter in $y$-axis and greater slope uncertainty. The same should also be true for the opposite pole orientation at $I=180^\circ$, which has a comparably-sized slope uncertainty.
The slope value for the lightcurves with $r_h>4.3$ au also deviated substantially from $\beta_L$. Notably, this was the lightcurve that covered the lowest range of phase angles ($0.2-12.6^\circ$) which likely impacted the straight line fit.

This highlights the challenge of fitting phase functions to photometric time series without lightcurve correction. 
When the lightcurve scatter is reduced - i.e. when we correctly account for the rotational brightness modulation of each lightcurve point and remove its corresponding amplitude from the reduced lightcurve - the slope value is consistently close to the true value. This was demonstrated in Section \ref{params}, in which the mean difference between $\beta_L$ and the corrected $\beta_{corr}$ across all noiseless lightcurves was reduced to 0.9 per cent. As shown in Figure \ref{fig:all_betas}, this is true on average even when the rotational variation is partially obscured by photon noise.  

To improve the reliability of literature phase function values, where possible these values should be corrected for rotation.
For dense lightcurves, this is typically in the form of rotational averaging, or correction using a convex model. 
It was recently demonstrated that removing rotational modulation had a significant impact on the phase function fit to JFC 162P/Siding Spring, which increased by 9 per cent from $0.0468\pm0.0001$ mag/$^\circ$ to $0.051\pm0.002$ mag/$^\circ$ \citep{donaldson_characterizing_2023}.
For populations with large amounts of available sparse photometry such as asteroids, methods are still being developed to fit reliable phase curves \citep[e.g.][]{martikainen_asteroid_2021, wilawer_asteroid_2022}. 
As identified by \cite{dobson_phase_2023} and \cite{robinson_main-belt_2024}, there are non-negligible differences between phase function fits to sparse lightcurves based on whether any attempt to account for the rotation is included.
Currently there are only around 20 JFCs for which it has been possible to estimate the nucleus phase function \citep[][and references therein]{knight_physical_2023}. 
Our analysis demonstrates that sparse LSST photometry alone is sufficient to accurately measure the rotationally-corrected phase function. 
This should be possible for many objects, leading to a vast improvement in our population statistics and a meaningful way to compare JFCs with other populations, to trace potential origins, evolution and dynamical links \citep[e.g.][]{2018kokotanekova}.

\section{Summary}

We generated synthetic JFC nucleus lightcurves for objects on a 67P orbit observed with an LSST-like cadence. 
Each synthetic lightcurve tested a different variant of nucleus characteristics, including rotation period, size, shape, obliquity and activity distance. 
In every case it was assumed that the comet was inactive during the observations, and that the nucleus was in a stable configuration with a constant rotation period.
We processed the sparse-in-time lightcurve points using convex lightcurve inversion to determine whether the input model properties could be reproduced.
The key findings and conclusions are summarised as follows:
\begin{enumerate}
    \item LSST will provide unprecedented coverage of the orbits of observable JFCs, detecting km-sized nuclei at large heliocentric distances with sufficient quality for lightcurve analysis. 
    Due to signal-to-noise limitations, we anticipate that such analysis will be possible for a few hundred objects with effective radii $\sim1$ km and larger.
    \item For every lightcurve tested, the best sidereal period identified by lightcurve inversion was consistent with the model input period. The input period was identifiable even with the addition of photon noise. This is extremely promising for accurately determining rotation periods for many JFCs following the ten-year operational period of LSST.
    
    \item The range of observing geometries covered by LSST observations for a typical low-inclination JFC is sufficient to constrain the nucleus pole orientation to within two possible directions. This will allow us to measure pole obliquities for more JFCs than is currently feasible.
    \item  The properties of the convex shapes determined from lightcurve 
    inversion of sparse survey data are comparable with those found by dense-in-time targeted monitoring campaigns. 
    
    \item It is possible to identify the correct rotation period and approximate pole orientations from lightcurve points spanning a three-year period.
    
    \item Throughout the survey duration, many JFCs will be observed at a larger range of phase angles than has been realistically attainable using ground-based facilities to date. In all test cases it was possible to rotationally-correct the lightcurves using the best-fit convex model, and recover a phase function value close to the model input value. We suggest that this becomes the standard practise for quoting JFC phase functions where possible to reduce uncertainty in the slope estimates from inherent lightcurve scatter. This will allow for more valid comparisons to be made between JFCs and other populations.

    \item Incorporating dense-in-time nucleus lightcurves with the sparse survey photometry measurably improves the accuracy of the period, pole and phase function estimates.

    \item Cometary activity is unpredictable and JFCs can exhibit activity at large heliocentric distances. Lightcurve points should carefully checked for signatures of sublimation-induced brightening, ideally automatically, prior to inclusion in any rotational lightcurve analysis.

\end{enumerate}

\begin{acknowledgments}

ACKNOWLEDGEMENTS

We thank the anonymous reviewers for their comments which helped improve the clarity of this work.
A.D. acknowledges support from the UK Science and Technology Facilities Council and ESO's SSDF 21/22 Student funding program. R.K. acknowledges the support from ``L’Oreal UNESCO For Women in Science" National program for Bulgaria.
This work has also benefited by the International Space Science Institute (ISSI) in Bern, through ISSI International Team project 504 ``The Life Cycle of Comets". 
For the purpose of open access, the author has applied a Creative Commons Attribution (CC BY) licence to any Author Accepted Manuscript version arising from this submission.
\end{acknowledgments}

\clearpage





\appendix
\section{Relevant plots for nucleus parameter variants}

\restartappendixnumbering
         
\begin{figure}[h]
    \centering
    \includegraphics[width=0.5\textwidth]{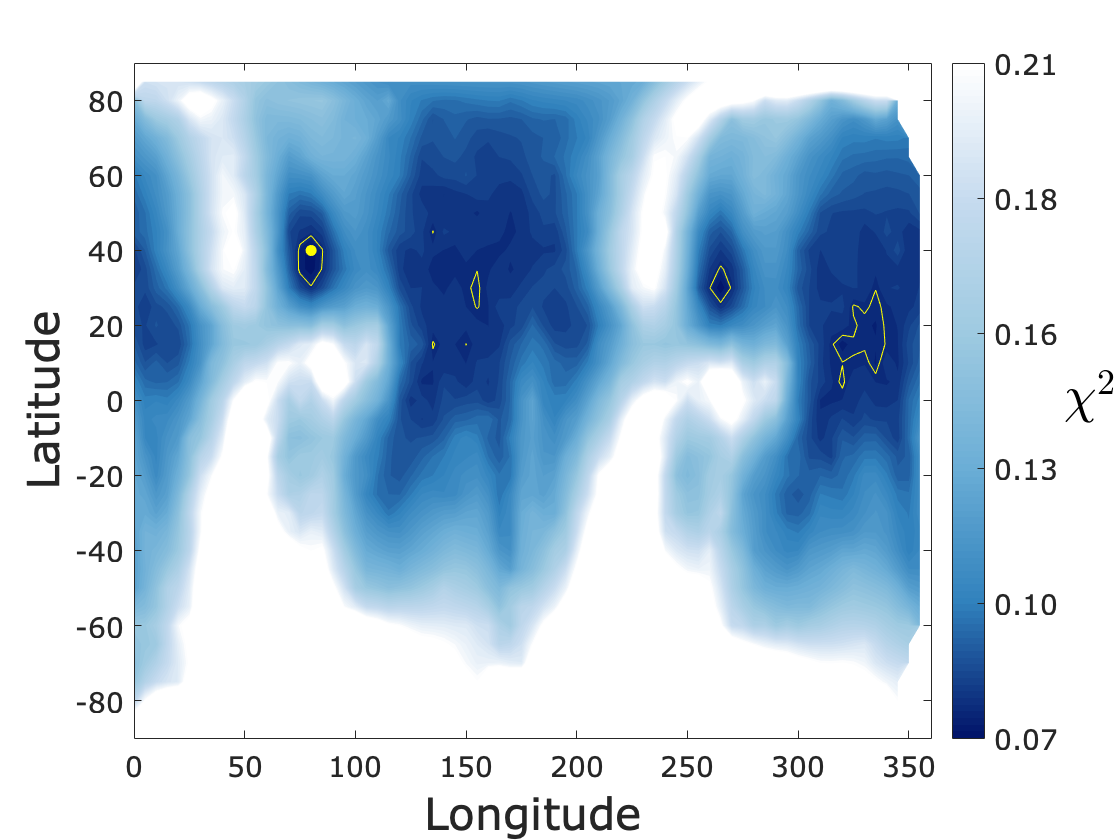}
    \caption{Distribution of $\chi^2$ values for tested pole orientations using a synthetic lightcurve generated with model input rotation period $P_i=23.9344696$h. The yellow lines enclose regions where the $\chi^2$ values lie within 10 per cent of the minimum $\chi^2$ value, which is located at the position of the yellow circle.}
    \label{fig:chi24}
\end{figure}

\begin{figure*}
\centering
\noindent
\begin{tabular}{c c}
       \includegraphics[trim={10cm 4cm 10cm 1cm},clip,width=0.5\textwidth]{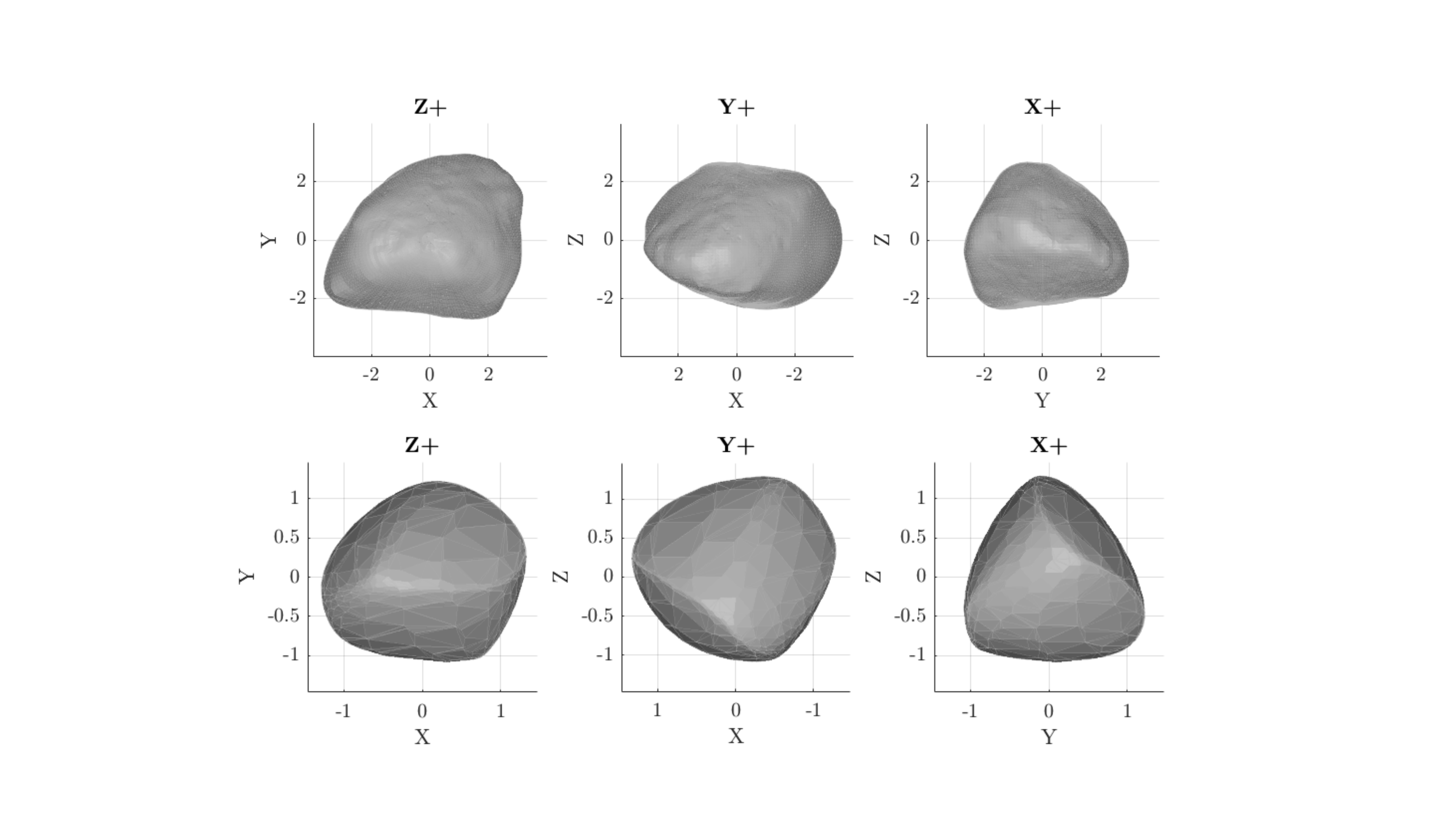} &
            \includegraphics[trim={10cm 2cm 10cm 1cm},clip,width=0.48\textwidth]{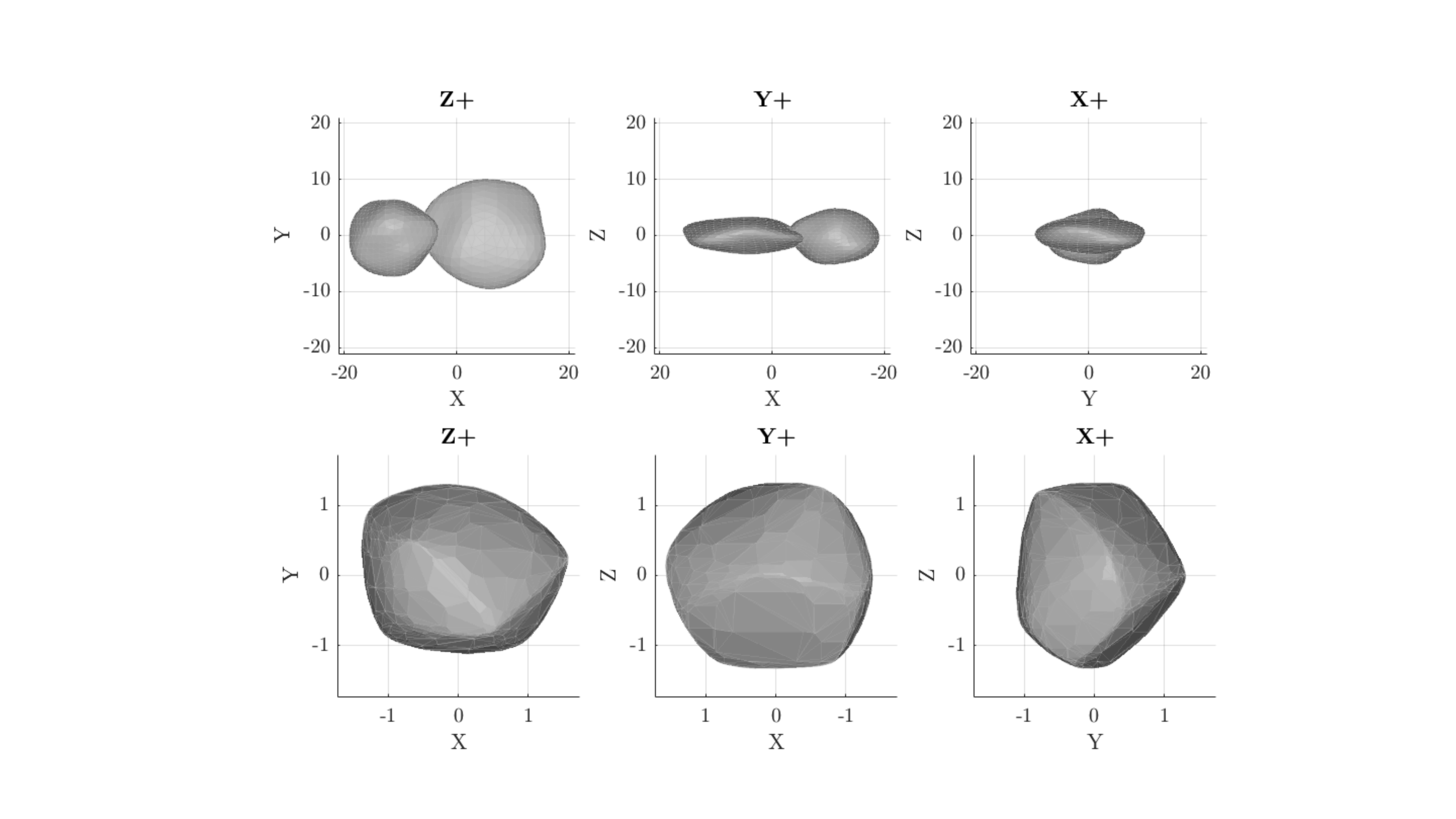}\\
            (a) & (b) \\
        \multicolumn{2}{c}{\includegraphics[trim={10cm 2cm 10cm 1cm},clip,width=0.5\textwidth]{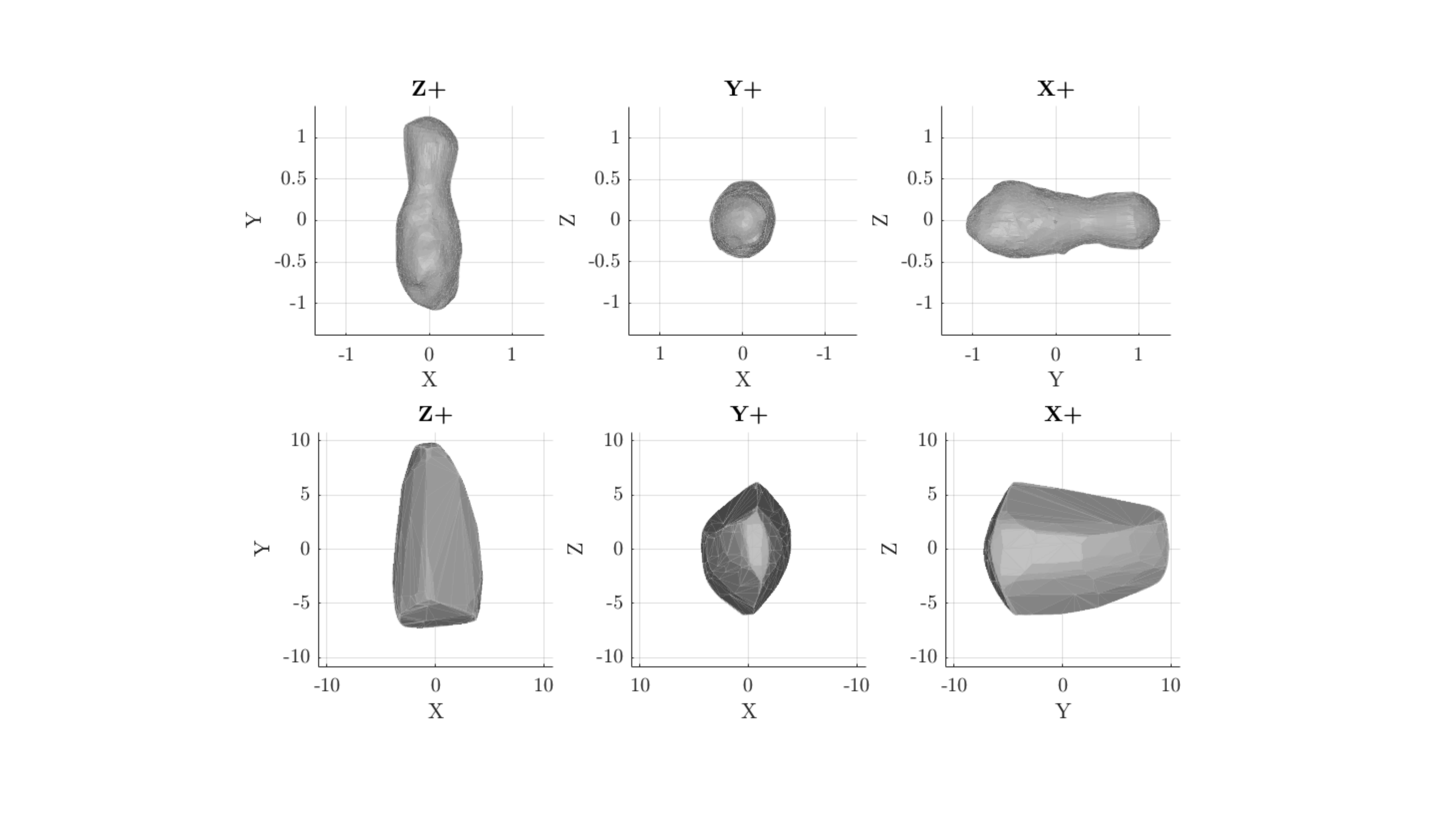}} \\
        \multicolumn{2}{c}{(c)} \\
\end{tabular}
\caption{A comparison of (upper) the representative shape model used to create the synthetic lightcurves and (lower) the best-fit shape model resulting from the lightcurve inversion analysis for a) 9P/Tempel 1; b) Arrokoth, and c) 103P/Hartley 2.}
    \label{fig:shapes}
\end{figure*}











\bibliography{bibliography}
\bibliographystyle{aasjournal}



\end{document}